\documentstyle[12pt,axodraw]{article}

\newcommand{\re}{\mbox{$\rm e$}}
\newcommand{\ri}{\mbox{$\rm i$}}

\newcommand{\sd}{\mbox{$\Sigma^\dagger$}}

\newcommand{\xid}{\mbox{$\xi^\dagger$}}

\newcommand{\bfm}[1]{\mbox{\boldmath$#1$}}
\newcommand{\ratio}[2]{\mbox{$#1\over#2$}}
\newcommand{\tr}{\mbox{$\,\rm Tr$}}

\begin{document}
\baselineskip=17pt
\parskip=5pt

\begin{titlepage}
\vfill

\hskip 4in {ISU-HET-98-2}

\hskip 4in {August 1998}
\vspace{1 in}
\begin{center}
{\large \bf Chiral Perturbation Theory for $|\Delta\bfm{I}|=3/2$  
Hyperon Decays}\\

\vspace{1 in}
{\bf  A.~Abd~El-Hady$^{(a,b)}$}, {\bf Jusak~Tandean$^{(a)}$}  
{\bf  and G.~Valencia$^{(a)}$}\\
{\it  $^{(a)}$ Department of Physics and Astronomy,
               Iowa State University,
               Ames IA 50011}\\
{\it  $^{(b)}$ Physics Department, 
               Zagazig University, 
               Zagazig, Egypt}\\
\vspace{1 in}
\end{center}
\begin{abstract}
 
We study the  $|\Delta\bfm{I}|=3/2$  amplitudes of hyperon  
non-leptonic decays of the form  $\;B\rightarrow B^\prime \pi\;$  
in the context of chiral perturbation theory. The lowest-order 
predictions are determined in terms of only one unknown parameter 
and are consistent within errors with current data. We investigate 
the theoretical uncertainty of these predictions by calculating the 
leading non-analytic corrections. We also present an estimate for 
the size of the S-wave $\Lambda$ and $\Xi$ decays which vanish at 
leading order. We find that the corrections to the lowest-order 
predictions are within the expectations of naive power counting 
and, therefore, that this picture can be tested more accurately 
with improved measurements.

\end{abstract}

\end{titlepage}

\clearpage

\section{Introduction}

Several papers have been devoted to the study of hyperon 
non-leptonic decays within the framework of chiral perturbation   
theory  ($\chi$PT)~\cite{bijnens,jenkins,hungary,springer,geocar}. 
These papers have dealt exclusively with the dominant  
$\;|\Delta\bfm{I}| =1/2\;$  transitions and have had mixed 
results.  
At leading order in  $\chi$PT,  these amplitudes can be  
parameterized in terms of two constants, and one-loop calculations  
have been carried out for the leading non-analytic terms. 
Whereas the S-waves  appear to be under control,  
the P-waves are not.

It has recently been pointed out that to leading order in  
$\chi$PT the  $\;|\Delta\bfm{I}|=3/2\;$  amplitudes of hyperon 
non-leptonic decays can be described in terms of only one 
parameter~\cite{heval}.   
In view of the situation in the $\;|\Delta\bfm{I}|=1/2\;$  sector,  
it is instructive to carry out a one-loop calculation in  
the  $\;|\Delta\bfm{I}|=3/2\;$  sector. 
This calculation of the corrections of  
${\cal O}(m_s^{}\ln{m_s^{}})$ to the leading amplitudes allows us 
to assess the reliability of the leading-order predictions.          
Two of the S-wave amplitudes (those for $\Lambda$ and $\Xi$ decays) 
vanish at leading order and are still zero after the 
non-analytic terms of ${\cal O}(m_s^{}\ln{m_s^{}})$ are included. 
We estimate the size of these two amplitudes by looking at some 
non-vanishing contributions to them.

This paper is organized as follows. 
In Section~2 we review the basic chiral Lagrangian for heavy 
baryons that we use for our calculation.   
In Section~3 we present detailed results 
of our calculation of the leading non-analytic corrections. 
In Section~4 we discuss two types of terms that can contribute to   
the S-wave amplitudes for  $\Lambda$ and $\Xi$ decays. 
In Section~5 we extract the experimental values of the 
$\;|\Delta\bfm{I}|=3/2\;$  amplitudes for the hyperon non-leptonic 
decays using the latest available information on decay rates and   
asymmetry parameters from the Particle Data Book~\cite{pdb}.  
We find slightly different numbers from the ones 
published in the Particle Data Book in 1982~\cite{overseth},  
but with similar and very large errors.  
We believe that there is an opportunity to improve at least 
some of these measurements with the large numbers of hyperon  
non-leptonic decays identified in two current experiments. 
Fermilab experiment E871, which searches for CP violation in 
hyperon decays, could improve the measurements of the decays of  
$\Lambda$ and $\Xi$ with one charged pion, and 
the KTeV experiment, which studies CP violation 
in  $\;K\rightarrow \pi\pi\;$  as well as rare kaon decays 
at Fermilab, could improve the measurements of the decays of  
$\Lambda$  and  $\Xi$  with a charged or neutral pion.   
Finally, in Section~6 we discuss our results.

\section{Chiral Lagrangian}

The chiral Lagrangian that describes the interactions of 
mesons and baryons is written down with the usual 
building blocks~\cite{georgi,dogoho}: 
the pseudo-Goldstone boson fields in the form of the matrix 
$\;\Sigma=\re^{{\rm i}\phi/f} ,\;$   where  $f$  is 
the pion-decay constant in the chiral-symmetry limit  and  
\begin{eqnarray}   
\phi  \;=\;  \sqrt{2}   
\left( \begin{array}{ccc}   \displaystyle  
\ratio{1}{\sqrt{2}}\, \pi^0+\ratio{1}{\sqrt{6}}\, \eta_8^{}  &    
\pi^+  &  K^+   \\ \\   
\pi^-  &  
\ratio{-1}{\sqrt{2}}\,\pi^0+\ratio{1}{\sqrt{6}}\,\eta_8^{}  &    
K^0  \\ \\   
K^-  &  \bar{K}^0  &  \ratio{-2}{\sqrt{6}}\, \eta_8^{}  
\end{array} \right) \;;     
\end{eqnarray}   
the octet baryons in the matrix 
\begin{eqnarray}   
B  \;=\;
\left( \begin{array}{ccc}   \displaystyle  
\ratio{1}{\sqrt{2}}\,\Sigma^0+\ratio{1}{\sqrt{6}}\,\Lambda  &    
\Sigma^+  &  p  \\ \\   
\Sigma^-  &  
\ratio{-1}{\sqrt{2}}\,\Sigma^0+\ratio{1}{\sqrt{6}}\,\Lambda  &    
n  \\ \\   
\Xi^-  &  \Xi^0  &  \ratio{-2}{\sqrt{6}}\, \Lambda  
\end{array} \right)   \;;     
\end{eqnarray}   
and the  spin-$\ratio{3}{2}$ decuplet baryons. Here we follow 
Jenkins and Manohar~\cite{hungary} and include the baryon-decuplet   
fields explicitly in the Lagrangian.   
As they argue, the mass splitting between the octet and decuplet 
baryons is small compared to the scale of chiral-symmetry breaking,   
and this enhances the effects of the decuplet on the low-energy 
theory. The decuplet baryons are described by a Rarita-Schwinger field  
$T^\mu_{abc}$, which satisfies the constraint  
$\;\gamma_\mu^{}T^\mu_{abc}=0 \;$   and is completely symmetric in 
its SU(3) indices,  $a,b,c$~\cite{hungary}.   
Its components are (with the Lorentz index suppressed)     
\begin{eqnarray}  
\begin{array}{c} \displaystyle      
T_{111}^{}  \;=\;  \Delta^{++}   \;, \hspace{2em}   
T_{112}^{}  \;=\;  \ratio{1}{\sqrt{3}}\, \Delta^+   \;, \hspace{2em}    
T_{122}^{}  \;=\;  \ratio{1}{\sqrt{3}}\, \Delta^0   \;, \hspace{2em}    
T_{222}^{}  \;=\;  \Delta^-   \;,   
\vspace{2ex} \\   \displaystyle       
T_{113}^{}  \;=\;  \ratio{1}{\sqrt{3}}\, \Sigma^{*+}   \;, \hspace{2em}    
T_{123}^{}  \;=\;  \ratio{1}{\sqrt{6}}\, \Sigma^{*0}   \;, \hspace{2em}    
T_{223}^{}  \;=\;  \ratio{1}{\sqrt{3}}\, \Sigma^{*-}   \;, 
\vspace{2ex} \\   \displaystyle       
T_{133}^{}  \;=\;  \ratio{1}{\sqrt{3}}\, \Xi^{*0}   \;, \hspace{2em}    
T_{233}^{}  \;=\;  \ratio{1}{\sqrt{3}}\, \Xi^{*-}   \;, \hspace{2em}    
T_{333}^{}  \;=\;  \Omega^-   \;. 
\end{array}    
\end{eqnarray}      
Under chiral SU(3$)_{\rm L}^{}\times~$SU(3$)_{\rm R}^{}$,   
these fields transform as   
\begin{eqnarray}    
\Sigma  \;\rightarrow\;  L \Sigma R^\dagger   \;, \hspace{2em}   
B  \;\rightarrow\;  U B U^\dagger             \;, \hspace{2em}   
T_{abc}^\mu  \;\rightarrow\;  
U_{ad}^{} U_{be}^{} U_{cf}^{} T_{def}^\mu    \;,    
\end{eqnarray}      
where  $\;L,R\in\,$SU$(3)_{\rm L,R}^{}$  and the matrix  $U$  is 
implicitly defined by the transformation  
\begin{eqnarray}    
\xi  \;\equiv\;  \re^{{\rm i}\phi/(2f)}  
\;\rightarrow\;  L\xi U^\dagger \;=\;  U\xi R^\dagger   \;.    
\end{eqnarray}      
We use the heavy-baryon formalism of Jenkins and  
Manohar~\cite{manjen}  where the effective Lagrangian is 
written in terms of velocity-dependent 
baryon fields, related to the ordinary baryon fields by   
the transformation~\cite{geor}     
\begin{eqnarray}    
B_v^{}(x)  \;=\;  
e^{{\rm i}m_B^{} \not{\,v}\, v\cdot x} \, B(x)
\;, \hspace{2em}   
T_v^\mu(x)  \;=\;  
e^{{\rm i}m_B^{} \not{\,v}\, v\cdot x}  \, T^\mu(x)  \;,   
\end{eqnarray}      
where  $m_B^{}$  is the baryon-octet mass in the chiral-symmetry limit.

The leading-order (in the derivative expansion), chiral- and 
parity-invariant Lagrangian that describes the strong interactions 
of the pseudoscalar-meson and baryon octets as well as the baryon  
decuplet is given  by~\cite{manjen,hungary}
\begin{eqnarray}   \label{L1strong}
{\cal L}^{\rm s}  &\!\!\!=&\!\!\!      
\ratio{1}{4} f^2\,
 \tr \!\left( \partial^\mu \Sigma^\dagger\, \partial_\mu \Sigma \right) 
\;+\;  
\tr \!\left( \bar{B}_v^{}\, \ri v\cdot {\cal D} B_v^{} \right)    
\nonumber \\ && \!\!\! \!\!\!   
+\; 
2 D\, \tr \!\left( \bar{B}_v^{}\, S_v^\mu 
 \left\{ {\cal A}_\mu^{}\,,\, B_v^{} \right\} \right)     
+ 2 F\, \tr \!\left( \bar{B}_v^{}\, S_v^\mu\,    
 \left[ {\cal A}_\mu^{}\,,\, B_v^{} \right] \right)   
\nonumber \\ && \!\!\! \!\!\!   
-\;    
\bar{T}_v^\mu\, \ri v\cdot {\cal D} T_{v\mu}^{}  
+ \Delta m\, \bar{T}_v^\mu T_{v\mu}^{}  
+ {\cal C} \left( \bar{T}_v^\mu {\cal A}_\mu^{} B_v^{} 
                   + \bar{B}_v^{} {\cal A}_\mu^{} T_v^\mu \right)    
+ 2{\cal H}\; \bar{T}_v^\mu\, S_v^{}\cdot{\cal A}\, T_{v\mu}^{}   \;,       
\end{eqnarray}      
where  $\Delta m$  denotes the mass difference between the decuplet 
and octet baryons in the chiral-symmetry limit, $S_v^\mu$ is the 
velocity dependent spin operator of Ref.~\cite{manjen},    
\begin{eqnarray}   
{\cal V}_\mu^{}  \;=\;  
\ratio{1}{2} 
\!\left( \xi\, \partial_\mu^{}\xid+\xid\, \partial_\mu^{}\xi  \right)   
\;, \hspace{2em}   
{\cal A}_\mu^{}  \;=\;  
\ratio{{\rm i}}{2} 
\!\left( \xi\, \partial_\mu^{}\xid-\xid\,\partial_\mu^{}\xi\right)   \;,   
\end{eqnarray}   
and 
\begin{equation}
{\cal D}^\mu B_v^{}  \;=\;    
\partial^\mu B_v^{} + [{\cal V}^\mu,B_v^{}]  \;,
\end{equation}   
\begin{eqnarray}    
{\cal D}^\mu \bigl( T_v^\nu \bigr) _{abc}^{}  \;=\;  
\partial^\mu \bigl( T_v^\nu \bigr) _{abc}^{} 
+ \ri{\cal V}_{ad}^\mu \bigl( T_v^\nu \bigr) _{dbc}^{}    
+ \ri{\cal V}_{bd}^\mu \bigl( T_v^\nu \bigr) _{adc}^{} 
+ \ri{\cal V}_{cd}^\mu \bigl( T_v^\nu \bigr) _{abd}^{}   \;,      
\end{eqnarray}      
\begin{eqnarray}    
\bar{T}_v^\mu {\cal A}_\mu^{} B_v^{} 
+ \bar{B}_v^{} {\cal A}_\mu^{} T_v^\mu  
\;=\;   
\epsilon_{abc}^{}\, \bigl( \bar{T}_v^\mu \bigr) _{cde}^{} 
\bigl( {\cal A}_\mu^{} \bigr) _{eb}^{} B_{da}^{} 
+ \epsilon_{abc}^{}\, \bar{B}_{ad}^{} 
 \bigl( {\cal A}_\mu^{} \bigr) _{be}^{} 
 \bigl( {T}_v^\mu \bigr) _{cde}^{}   \;.   
\end{eqnarray}      
Explicit breaking of chiral symmetry, to leading order in 
the mass of the strange quark and in the limit  
$\;m_u^{} = m_d^{} =0,\;$  is introduced via  
the  Lagrangian~\cite{jenmass}
\begin{eqnarray}   \label{Lmstrong}
{\cal L}_{m_q^{}}^{\rm s}  &\!\!\!=&\!\!\!    
a\, \tr \!\left( M\Sigma^\dagger+\Sigma M^\dagger \right) 
\nonumber \\ && \!\!\! \!\!\!   
+\; 
b_D^{}\,   
 \tr \!\left( \bar{B}_v^{} 
           \left\{ \xid M\xid+\xi M^\dagger\xi\,,\, B_v^{} \right\} \right)   
+ b_F^{}\,    
 \tr \!\left( \bar{B}_v^{}  
           \left[ \xid M\xid+\xi M^\dagger\xi\,,\, B_v^{}  \right] \right)   
\nonumber \\ && \!\!\! \!\!\!   
+\; 
\sigma\, \tr \!\left( M\Sigma^\dagger+\Sigma M^\dagger \right) 
 \tr \!\left( \bar{B}_v^{} B_v^{}  \right)  
\nonumber \\ && \!\!\! \!\!\!   
+\; 
c\, \bar{T}_v^\mu \left( \xid M\xid+\xi M^\dagger\xi \right) T_{v\mu}^{}      
- \tilde{\sigma}\, \tr \!\left( M\Sigma^\dagger+\Sigma M^\dagger \right)  \,  
 \bar{T}_v^\mu T_{v\mu}^{}   \;,          
\end{eqnarray}   
where $\;M={\rm diag}(0,0,m_s^{}).\;$  
In this limit, the pion is massless and the  $\eta_8^{}$  mass is 
related to the kaon mass by  
$\;m_{\eta_8^{}}^2=\ratio{4}{3}\, m_K^2 .\;$    
Furthermore, mass splittings within the baryon octet and decuplet 
occur to linear order in  $m_s^{}$.

Within the standard model,  the  $\;|\Delta S|=1$,  
$\;|\Delta\bfm{I}|=3/2\;$  transitions are induced by an effective  
Hamiltonian that transforms as  $(27_{\rm L}^{},1_{\rm R}^{})$  
under chiral rotations:   
\begin{equation}
{\cal H}_{\rm eff}^{(27_{\rm L}^{},1_{\rm R}^{})}  \;=\; 
{G_{\rm F}^{} \over \sqrt{2}} V^*_{ud} V_{us}^{} 
\left( {c_1+c_2 \over 3} \right)
{\cal O}^{(27_{\rm L}^{},1_{\rm R}^{})}_{|\Delta I|=3/2} 
\;+\;  {\rm h.c.}
\label{weaktsl}
\end{equation}
The four-quark operator  
$\;{\cal O}^{(27_{\rm L}^{},1_{\rm R}^{})}_{|\Delta I|=3/2} = 
4 T_{jk,lm}^{}\, \bar{\psi}^j_{\rm L} \gamma_\mu^{} \psi^l_{\rm L}\,  
\bar{\psi}^k_{\rm L} \gamma^\mu \psi^m_{\rm L}\;$   
has a unique chiral realization in the baryon-octet sector 
at leading order in  $\chi$PT~\cite{heval}.   
Similarly, at leading-order in $\chi$PT,  there is only one 
operator with the required transformation properties involving 
two decuplet baryon fields, and there are no operators that involve 
one decuplet-baryon and one octet-baryon fields. 
The leading-order weak chiral Lagrangian is, thus, 
\begin{eqnarray}   
{\cal L}^{\rm w}  \;=\;     
\beta_{27}^{}\, 
T_{ij,kl}^{} \left( \xi\bar{B}_v^{} \xid \right) _{\!ki}^{} 
\left( \xi B_v^{} \xid \right) _{\!lj}^{}   
\,+\,  
\delta_{27}^{}\, 
T_{ij,kl}^{}\; \xi_{kd}^{} \xi_{bi}^\dagger\; 
\xi_{le}^{} \xi_{cj}^\dagger\; 
\bigl( \bar{T}_v^\mu \bigr) _{abc}^{} 
\bigl( T_{v\mu}^{} \bigr) _{ade}^{} 
\;+\;  {\rm h.c.}   
\label{loweak}
\end{eqnarray}      
The non-zero elements of $T_{ij,kl}^{}$ that project out the  
$\;|\Delta S|=1$,  $\;|\Delta\bfm{I}|=3/2\;$  Lagrangian are 
$\;T_{12,13}^{}=T_{21,13}^{}=T_{12,31}^{}=T_{21,31}^{}=1/2\;$ 
and  $\;T_{22,23}^{}=T_{22,32}^{}=-1/2\;$~\cite{georgi}.       
In order to simplify notation and to parallel the discussions for 
the  $\;|\Delta\bfm{I}|=1/2\;$  sector of 
Refs.~\cite{bijnens,jenkins}, unlike Ref.~\cite{heval}, we absorb  
the Fermi constant, the CKM angles and the Wilson coefficients
in  Eq.~(\ref{weaktsl}) into the constants $\beta_{27}^{}$  and  
$\delta_{27}^{}$  in  Eq.~(\ref{loweak}). The $\;|\Delta\bfm{I}|=1/2\;$ 
weak Lagrangian of Ref.~\cite{jenkins},  which transforms   
as  $(8_{\rm L}^{},1_{\rm R}^{})$, is given by   
\begin{eqnarray}   
{\cal L}^{\rm w,8}  \;=\;     
h_{D}^{}\, \tr 
\!\left( \bar{B}_v^{} \left\{ \xid h \xi\,,\,B_v^{} \right\} \right) 
\,+\,  
h_{F}^{}\, \tr   
\!\left( \bar{B}_v^{} \left[ \xid h \xi\,,\,B_v^{} \right]  \right) 
\,+\,  
h_{C}^{}\, \bar{T}_v^\mu\, \xid h \xi\, T_{v\mu}^{} 
\;+\;  {\rm h.c.}   \;,  
\label{loweakoc}
\end{eqnarray}      
where  $\;h_{23}^{}=1\;$  and all other elements of $h$ vanish.
Although this notation has become standard, for power counting arguments 
we will find it  convenient to assume an explicit factor of   
$\;G_{\rm F}^{} f_{\!\pi}^3 V_{ud}^{*} V_{us}^{}\;$ multiplying 
$\beta_{27}^{}$, $h_{D}^{}$ and $h_{F}^{}$.

For purely-mesonic  $\;|\Delta S|=1$,  $\;|\Delta\bfm{I}|=3/2\;$   
processes, the lowest-order weak Lagrangian can be written as    
\begin{eqnarray}   
{\cal L}^{\rm w}_\phi  \;=\;  
{G_{\rm F}^{} \over \sqrt{2}} 
f_{\!\pi}^4 V_{ud}^{*} V_{us}^{}\,g_{27}^{}\, 
T_{ij,kl}^{} \left( \partial^\mu \Sigma\,\sd \right) _{\!ki}^{}  
\left( \partial_\mu^{} \Sigma\,\sd \right) _{\!lj}^{}   
\;+\;  {\rm h.c.}  
\label{mesonwe}
\end{eqnarray}   
As defined in this expression, the constant $g_{27}^{}$ is  
expected to be of order one. Experimentally, it turns out that 
$\;g_{27}^{}\approx 0.16\;$  as extracted from an analysis of 
$\;K \rightarrow \pi \pi\;$  decays~\cite{cronin,phyrep,devlin}.  
For comparison, the analogous constant for   
$\;|\Delta\bfm{I}|=1/2\;$  transitions, $g_8^{}$, is measured 
to be  $\;g_8^{}\approx 5.1\;$ and is expected to be of order 
one when defined by the Lagrangian  
\begin{eqnarray}   
{\cal L}^{{\rm w},8}_\phi  \;=\;  
{G_{\rm F}^{} \over \sqrt{2}} 
f_{\!\pi}^4 V_{ud}^{*} V_{us}^{}\,g_8^{}\, 
\tr \!\left( h\, \partial_\mu^{} \Sigma\,  
            \partial^\mu \Sigma^\dagger \right)  
\;+\;  {\rm h.c.}   \;,   
\label{mesonweoc}  
\end{eqnarray}      
which transforms as  $(8_{\rm L}^{},1_{\rm R}^{})$   
under chiral symmetry.

\section{$|\Delta\bfm{I}|=3/2$ Amplitudes to 
${\cal O}(m_s^{}\ln{m_s^{}})$}

There are two terms in the amplitude for the decay   
$\;B\rightarrow B^\prime\pi,\;$  corresponding to S- and P-wave 
contributions.   
In our calculation we refer exclusively to the 
$\;|\Delta\bfm{I}|=3/2\;$  component of these amplitudes. 
We use the heavy-baryon approach and  
follow Ref.~\cite{jenkins} to write the amplitude in the form
\begin{eqnarray}        
\ri {\cal M}^{}_{B_{}\rightarrow B_{}'\pi}   \;=\;  
G_{\rm F}^{} m_{\pi}^2\, 
\bar{u}_{B_{}'}^{} \left( 
{\cal A}^{(\rm S)}_{B_{}^{}B_{}'\pi}   
+ 2 k\cdot S_v^{}\, {\cal A}^{(\rm P)}_{B_{}^{}B_{}'\pi} 
\right) u_{B_{}^{}}^{}   \;,  
\end{eqnarray}    
where  $k$  is the outgoing four-momentum of the pion.   
The  $\;|\Delta\bfm{I}|=3/2\;$  amplitudes satisfy 
the isospin relations  
\begin{eqnarray}   
\begin{array}{c}   \displaystyle        
{\cal M}^{}_{\Sigma^+\rightarrow n\pi^+}
- \sqrt{2}\, {\cal M}^{}_{\Sigma^+\rightarrow p\pi^0}
+ 2 {\cal M}^{}_{\Sigma^-\rightarrow n\pi^-}  \;=\;  0   \;,  
\vspace{3ex} \\   \displaystyle  
{\cal M}^{}_{\Lambda\rightarrow n\pi^0}
- \sqrt{2}\, {\cal M}^{}_{\Lambda\rightarrow p\pi^-}  \;=\;  0   \;,
\vspace{3ex} \\   \displaystyle    
{\cal M}^{}_{\Xi^0\rightarrow \Lambda\pi^0}
- \sqrt{2}\, {\cal M}^{}_{\Xi^-\rightarrow \Lambda\pi^-}  \;=\;  0   \;, 
\end{array}   
\end{eqnarray}   
\hfill \\   
and, therefore, only four of them are independent. 
We have chosen to present  $\;\Sigma^+\rightarrow n\pi^+,\;$  
$\Sigma^-\rightarrow n\pi^-,\;$  $\Lambda\rightarrow p\pi^-\;$   
and  $\;\Xi^-\rightarrow\Lambda\pi^-\;$  because these are 
the same ones used in the discussion of  
$\;|\Delta\bfm{I}|=1/2\;$  transitions~\cite{bijnens,jenkins}.

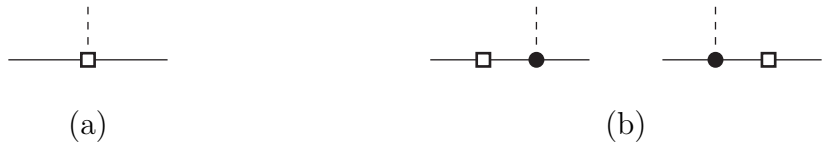
\begin{figure}[t]         
   \hspace*{\fill} 
\begin{picture}(60,70)(-30,-35)    
\Line(-30,0)(30,0) \DashLine(0,0)(0,20){3} 
\Text(0,-25)[c]{(a)}   
\SetWidth{1}   \BBoxc(0,0)(5,5)         
\end{picture}
   \hspace*{\fill} 
\begin{picture}(60,70)(-30,-35)
\Line(-30,0)(30,0)  
\Vertex(10,0){3} \DashLine(10,0)(10,20){3}    
\SetWidth{1}   \BBoxc(-10,0)(5,5) 
\end{picture}
\begin{picture}(20,70)(-10,-35)
\Text(0,-25)[c]{(b)}   
\end{picture}
\begin{picture}(60,70)(-30,-35)
\Line(-30,0)(30,0)  
\Vertex(-10,0){3} \DashLine(-10,0)(-10,20){3}    
\SetWidth{1}   \BBoxc(10,0)(5,5)          
\end{picture}
   \hspace*{\fill} 
\caption{\label{tree}%
Tree-level diagrams for (a) S-wave and (b) P-wave hyperon 
non-leptonic decays. 
Solid (dashed) lines denote baryon-octet (meson-octet) fields.   
A solid dot (hollow square) represents a strong (weak) vertex.
In all figures, the strong vertices are generated by  
${\cal L}^{\rm s}$  in Eq.~(\ref{L1strong}).  
Here the weak vertices come from  ${\cal L}^{\rm w}$  in  
Eq.~(\ref{loweak}).}
\end{figure}

We follow the notation of Jenkins~\cite{jenkins} to write the 
S- and P-wave decay amplitudes at the one-loop level in the form    
\vspace{1ex} 
\begin{eqnarray}   
{\cal A}^{\rm (S)}_{B_{}^{}B_{}'\pi}  \;=\;   
{1\over\sqrt{2}\, f_{\!\pi}^{}} \Biggl[  
\alpha^{\rm (S)}_{B_{}^{}B_{}'}     
\;+\;  
\Bigl( \bar{\beta}^{\rm (S)}_{B_{}^{}B_{}'}     
      - \bar{\lambda}^{}_{B_{}^{}B_{}'\pi} 
       \alpha^{\rm (S)}_{B_{}^{}B_{}'} \Bigl)      
{m_K^2\over 16\pi^2 f_{\!\pi}^2}\, \ln{m_K^2\over\mu^2}  
\Biggr]   \;,   
\label{defswave}
\end{eqnarray}    
\begin{eqnarray}   
{\cal A}^{\rm (P)}_{B_{}^{}B_{}'\pi}  \;=\;   
{1\over\sqrt{2}\, f_{\!\pi}^{}} \Biggl[  
\alpha^{\rm (P)}_{B_{}^{}B_{}'}     
\;+\;  
\Bigl( \bar{\beta}^{\rm (P)}_{B_{}^{}B_{}'}     
        - \bar{\lambda}^{}_{B_{}^{}B_{}'\pi} 
         \alpha^{\rm (P)}_{B_{}^{}B_{}'} \Bigl)      
 {m_K^2\over 16\pi^2 f_{\!\pi}^2}\, \ln{m_K^2\over\mu^2}  
\;+\;  
\gamma_{B_{}^{}B_{}'}^{} \alpha^{\rm (P)}_{B_{}^{}B_{}'}   
\Biggr]   \;,
\label{defpwave}   
\end{eqnarray}    
\hfill \\  
where  $\;f_{\!\pi}^{}\approx 92.4\, \rm MeV\;$  is the physical 
pion-decay constant;  
$\;\alpha^{}_{B_{}^{}B_{}'}\;$  and  
$\; \bar{\beta}^{}_{B_{}^{}B_{}'} =     
   \beta^{}_{B_{}^{}B_{}'} + \beta'_{B_{}^{}B_{}'}\;$    
represent contributions from tree-level and one-loop diagrams, 
respectively, shown in Figures~\ref{tree},  \ref{s-wave,loop},  
\ref{p-wave,loop},  and~\ref{g27,loop};  
$\lambda^{}_{B_{}^{}B_{}'\pi}$  arises from  
baryon and pion wave-function renormalization as well as 
the renormalization of the pion-decay constant;  
and  $\gamma^{}_{B_{}^{}B_{}'}$  results from the one-loop 
corrections to the propagator in the P-wave diagrams.\footnote{%
A similar contribution to P-wave amplitudes in 
the  $\;|\Delta\bfm{I}|=1/2\;$  sector was not included in 
the calculation of Jenkins~\cite{jenkins} and  was pointed out by  
Springer~\cite{springer}.}        
One-loop decay graphs involving only octet baryons contribute to 
$\beta^{}_{B_{}^{}B_{}'}$,  whereas those with internal   
decuplet-baryon lines yield  $\beta'_{B_{}^{}B_{}'}$.

At leading order in $\chi$PT, ${\cal O}(1)$, there are 
contributions to the amplitudes from the tree-level Lagrangian 
in  Eq.~(\ref{loweak}).  
They arise from the diagrams displayed in Figure~\ref{tree} and 
are given by
\vspace{1ex} 
\begin{eqnarray}        
\alpha^{(\rm S)}_{\Sigma^+ n}  \;=\;  -\ratio{3}{2} \beta_{27}^{}     
\;, \hspace{2em}         
\alpha^{(\rm S)}_{\Sigma^- n}  \;=\;  \beta_{27}^{}   
\;, \hspace{2em}         
\alpha^{(\rm S)}_{\Lambda p}  \;=\;  0   \;, \hspace{2em}         
\alpha^{(\rm S)}_{\Xi^-\Lambda}  \;=\;  0   \;,           
\end{eqnarray}    
\begin{eqnarray}        
\begin{array}{rlrl}   \displaystyle  
\vspace{-1ex} \\   
\alpha^{(\rm P)}_{\Sigma^+ n}  &\!\!=\;   \displaystyle  
\Bigl( -\ratio{1}{2} D - \ratio{3}{2} F \Bigr) 
{\beta_{27}^{}\over m_\Sigma^{}-m_N^{}}   
\;, \hspace{3em} &   
\alpha^{(\rm P)}_{\Sigma^- n}  &\!\!=\;   \displaystyle  
F\, {\beta_{27}^{}\over m_\Sigma^{}-m_N^{}}   \;,     
\vspace{3ex} \\   \displaystyle    
\alpha^{(\rm P)}_{\Lambda p}     &\!\!=\;   \displaystyle      
\ratio{1}{\sqrt{6}} D\, {\beta_{27}^{}\over m_\Sigma^{}-m_N^{}}   
\;, &   
\alpha^{(\rm P)}_{\Xi^-\Lambda}  &\!\!=\;   \displaystyle   
-\ratio{1}{\sqrt{6}} D\, {\beta_{27}^{}\over m_\Xi^{}-m_\Sigma^{}}   \;, 
\end{array}  
\end{eqnarray}    
where the strong vertices in the P-wave graphs are from 
${\cal L}^{\rm s}$  in Eq.~(\ref{L1strong}).
These results correspond to those obtained in Ref.~\cite{heval} without 
the heavy-baryon formalism.\footnote{%
The coupling  $\beta_{27}^{}$  is proportional to the constant 
$b_{27}^{}$  in Ref.~\cite{heval}}  
\begin{figure}[t]         
   \hspace*{\fill} 
\begin{picture}(60,70)(-30,-30)    
\Line(-30,0)(30,0) \DashLine(0,0)(0,20){3} 
\DashCArc(0,-10)(10,0,360){3}        
\SetWidth{1}   \BBoxc(0,0)(5,5)         
\end{picture}
   \hspace*{\fill} 
\begin{picture}(80,70)(-40,-30)
\Line(-40,0)(40,0)
\DashCArc(0,0)(20,180,360){4} 
\Vertex(20,0){3} \DashLine(20,0)(20,20){3}    
\SetWidth{1}   \BBoxc(-20,0)(5,5) 
\end{picture}
\begin{picture}(20,70)(-10,-30)
\end{picture}
\begin{picture}(80,70)(-40,-30)
\Line(-40,0)(40,0) 
\DashCArc(0,0)(20,180,360){4} 
\Vertex(-20,0){3} \DashLine(-20,0)(-20,20){3}    
\SetWidth{1}   \BBoxc(20,0)(5,5)          
\end{picture}
   \hspace*{\fill} 
\\    
   \hspace*{\fill} 
\begin{picture}(80,60)(-40,-30)
\Line(-40,0)(40,0) 
\DashCArc(0,0)(20,180,360){4} 
\Vertex(-20,0){3} \Vertex(20,0){3} 
\DashLine(0,0)(0,20){3}    
\SetWidth{1}   \BBoxc(0,0)(5,5)             
\end{picture}
   \hspace*{\fill} 
\begin{picture}(80,60)(-40,-30)
\Line(-40,0)(-20,0) 
\Line(-20,1.5)(20,1.5) \Line(-20,-1.5)(20,-1.5) 
\Line(20,0)(40,0) 
\DashCArc(0,0)(20,180,360){4} 
\Vertex(-20,0){3} \Vertex(20,0){3} 
\DashLine(0,0)(0,20){3}    
\SetWidth{1}   \BBoxc(0,0)(5,5)             
\end{picture}
   \hspace*{\fill} 
\caption{\label{s-wave,loop}%
One-loop diagrams contributing to S-wave hyperon non-leptonic 
decay amplitudes, with weak vertices from  ${\cal L}^{\rm w}$  
in  Eq.~(\ref{loweak}).   
Double (single) solid-lines represent baryon-decuplet 
(baryon-octet) fields.}  
\end{figure}
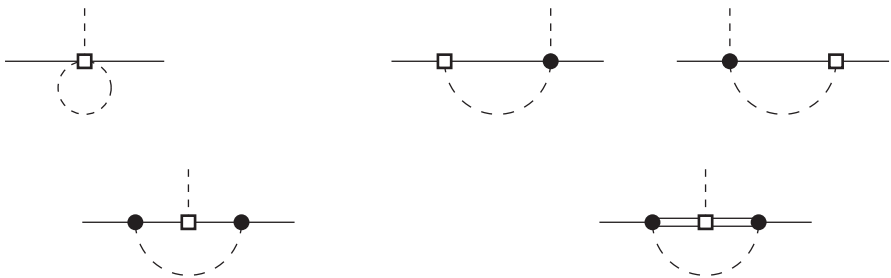             

At next order in $\chi$PT,  we will have amplitudes of 
${\cal O}(m_s^{})$ arising both from one-loop diagrams with 
lowest-order vertices and from counterterms. 
At present, there is not enough experimental input to determine 
the value of the counterterms. 
For this reason, we follow the approach that has been used for the   
$\;|\Delta\bfm{I}|=1/2\;$  amplitudes~\cite{bijnens,jenkins} and  
calculate only those terms of  ${\cal O}(m_s^{}\ln{m_s^{}})$.  
These terms are uniquely determined from the one-loop amplitudes   
because they cannot arise from local counterterm Lagrangians.   
$\chi$PT purists may argue that this is an incomplete calculation.   
However, our calculation will suffice to estimate the robustness 
of the leading-order predictions. 
With a complete calculation at next-to-leading order, it would be
possible to fit all the amplitudes,\footnote{We have not constructed 
a complete list of the operators that occur at next-to-leading order. 
However, we have verified that there are at least six new terms so that,
at  ${\cal O}(m_s^{})$, there are more unknown constants than there 
are measurements. This is analogous to what happens for 
$\;|\Delta\bfm{I}|=1/2\;$ amplitudes~\cite{borahol}.}   
but we feel  that there is nothing to be learned from that exercise given  
the large number of free parameters available.

\vfill   \newpage   
\begin{figure}[ht]         
   \hspace*{\fill} 
\begin{picture}(80,40)(-40,-10)    
\Line(-40,0)(-15,0) 
\DashCArc(-15,10)(10,-90,270){3}        
\Line(-15,0)(15,0) 
\Vertex(15,0){3} \DashLine(15,0)(15,20){3} 
\Line(15,0)(40,0) 
\SetWidth{1}   \BBoxc(-15,0)(5,5) 
\end{picture}
   \hspace*{\fill} 
\begin{picture}(80,40)(-40,-10)    
\Line(-40,0)(-15,0) \Vertex(-15,0){3} 
\DashLine(-15,0)(-15,20){3} 
\Line(-15,0)(15,0) 
\DashCArc(15,10)(10,-90,270){3}        
\Line(15,0)(40,0) 
\SetWidth{1}   \BBoxc(15,0)(5,5) 
\end{picture}
   \hspace*{\fill} 
\\    
   \hspace*{\fill} 
\begin{picture}(80,60)(-40,-30)    
\Line(-40,0)(-15,0) \Vertex(15,0){3} 
\Line(-15,0)(15,0) 
\DashLine(15,0)(15,20){3} 
\DashCArc(15,-10)(10,0,360){3}        
\Line(15,0)(40,0) 
\SetWidth{1}   \BBoxc(-15,0)(5,5) 
\end{picture}
   \hspace*{\fill} 
\begin{picture}(80,60)(-40,-30)    
\Line(-40,0)(-15,0) \Vertex(-15,0){3}     
\DashLine(-15,0)(-15,20){3}   
\DashCArc(-15,-10)(10,-180,180){3}        
\Line(-15,0)(15,0) \Line(15,0)(40,0) 
\SetWidth{1}   \BBoxc(15,0)(5,5)         
\end{picture}               
   \hspace*{\fill} 
\\    
   \hspace*{\fill} 
\begin{picture}(100,60)(-60,-30)
\Line(-60,0)(-40,0) \Line(-40,0)(-20,0)       
\Line(-20,0)(0,0) \Line(0,0)(20,0) \Line(20,0)(40,0) 
\DashCArc(0,0)(20,180,0){4} 
\Vertex(-20,0){3} \Vertex(0,0){3} \Vertex(20,0){3} 
\DashLine(0,0)(0,20){3} 
\SetWidth{1}   \BBoxc(-40,0)(5,5) 
\end{picture}
   \hspace*{\fill} 
\begin{picture}(100,60)(-40,-30)
\Line(-40,0)(-20,0) \Line(-20,0)(0,0) \Line(0,0)(20,0) 
\DashCArc(0,0)(20,180,0){4}        
\Vertex(-20,0){3} \Vertex(0,0){3} \Vertex(20,0){3}         
\DashLine(0,0)(0,20){3} 
\Line(20,0)(40,0) \Line(40,0)(60,0) 
\SetWidth{1}   \BBoxc(40,0)(5,5) 
\end{picture}
   \hspace*{\fill} 
\\  
   \hspace*{\fill} 
\begin{picture}(100,60)(-60,-30)
\Line(-60,0)(-40,0) \Line(-40,0)(-20,0)       
\Line(-20,1.5)(20,1.5) \Line(-20,-1.5)(20,-1.5) 
\Line(20,0)(40,0) 
\DashCArc(0,0)(20,180,0){4} 
\Vertex(-20,0){3} \Vertex(0,0){3} \Vertex(20,0){3} 
\DashLine(0,0)(0,20){3} 
\SetWidth{1}   \BBoxc(-40,0)(5,5) 
\end{picture}   
   \hspace*{\fill} 
\begin{picture}(100,60)(-40,-30)
\Line(-40,0)(-20,0) 
\Line(-20,1.5)(20,1.5) \Line(-20,-1.5)(20,-1.5) 
\DashCArc(0,0)(20,180,0){4}        
\Vertex(-20,0){3} \Vertex(0,0){3} \Vertex(20,0){3}         
\DashLine(0,0)(0,20){3} 
\Line(20,0)(40,0) \Line(40,0)(60,0) 
\SetWidth{1}   \BBoxc(40,0)(5,5) 
\end{picture}
   \hspace*{\fill} 
\\    
   \hspace*{\fill} 
\begin{picture}(100,60)(-60,-30)
\Line(-60,0)(-40,0) \Line(-40,0)(-20,0)       
\Line(-20,1.5)(0,1.5) \Line(-20,-1.5)(0,-1.5)   
\Line(0,0)(20,0) \Line(20,0)(40,0) 
\DashCArc(0,0)(20,180,0){4} 
\Vertex(-20,0){3} \Vertex(0,0){3} \Vertex(20,0){3} 
\DashLine(0,0)(0,20){3} 
\SetWidth{1}   \BBoxc(-40,0)(5,5) 
\end{picture}
   \hspace*{\fill} 
\begin{picture}(100,60)(-40,-30)
\Line(-40,0)(-20,0) 
\Line(-20,1.5)(0,1.5) \Line(-20,-1.5)(0,-1.5) 
\Line(0,0)(20,0)  \DashCArc(0,0)(20,180,0){4}        
\Vertex(-20,0){3} \Vertex(0,0){3} \Vertex(20,0){3}         
\DashLine(0,0)(0,20){3} 
\Line(20,0)(40,0) \Line(40,0)(60,0) 
\SetWidth{1}   \BBoxc(40,0)(5,5) 
\end{picture}
   \hspace*{\fill} 
\\  
   \hspace*{\fill} 
\begin{picture}(100,60)(-60,-30)
\Line(-60,0)(-40,0) \Line(-40,0)(-20,0)       
\Line(-20,0)(0,0) \Line(0,1.5)(20,1.5) \Line(0,-1.5)(20,-1.5) 
\Line(20,0)(40,0) 
\DashCArc(0,0)(20,180,0){4} 
\Vertex(-20,0){3} \Vertex(0,0){3} \Vertex(20,0){3} 
\DashLine(0,0)(0,20){3} 
\SetWidth{1}   \BBoxc(-40,0)(5,5) 
\end{picture}
   \hspace*{\fill} 
\begin{picture}(100,60)(-40,-30)
\Line(-40,0)(-20,0) \Line(-20,0)(0,0) 
\Line(0,1.5)(20,1.5) \Line(0,-1.5)(20,-1.5) 
\DashCArc(0,0)(20,180,0){4}        
\Vertex(-20,0){3} \Vertex(0,0){3} \Vertex(20,0){3}         
\DashLine(0,0)(0,20){3} 
\Line(20,0)(40,0) \Line(40,0)(60,0) 
\SetWidth{1}   \BBoxc(40,0)(5,5) 
\end{picture}
   \hspace*{\fill} 
\\  
   \hspace*{\fill} 
\begin{picture}(100,40)(-40,-10)
\Line(-40,0)(-20,0) \Line(-20,0)(0,0) \Line(0,0)(20,0) 
\DashCArc(0,0)(20,0,180){4}        
\Vertex(-20,0){3} \Vertex(20,0){3} \Vertex(40,0){3}         
\DashLine(40,0)(40,20){3} 
\Line(20,0)(40,0) \Line(40,0)(60,0) 
\SetWidth{1}   \BBoxc(0,0)(5,5) 
\end{picture}
   \hspace*{\fill} 
\begin{picture}(100,40)(-60,-10)
\Line(-60,0)(-40,0) 
\Line(-40,0)(-20,0)       
\DashLine(-40,0)(-40,20){3} 
\Line(-20,0)(0,0) \Line(0,0)(20,0) \Line(20,0)(40,0) 
\DashCArc(0,0)(20,0,180){4} 
\Vertex(-40,0){3} \Vertex(-20,0){3} \Vertex(20,0){3} 
\SetWidth{1}   \BBoxc(0,0)(5,5) 
\end{picture}
   \hspace*{\fill} 
\\    
   \hspace*{\fill} 
\begin{picture}(100,40)(-40,-10)
\Line(-40,0)(-20,0) 
\Line(-20,1.5)(0,1.5) \Line(-20,-1.5)(0,-1.5) 
\Line(0,1.5)(20,1.5)  \Line(0,-1.5)(20,-1.5)  
\DashCArc(0,0)(20,0,180){4}        
\Vertex(-20,0){3} \Vertex(20,0){3} \Vertex(40,0){3}         
\DashLine(40,0)(40,20){3} 
\Line(20,0)(40,0) \Line(40,0)(60,0) 
\SetWidth{1}   \BBoxc(0,0)(5,5) 
\end{picture}
   \hspace*{\fill} 
\begin{picture}(100,40)(-60,-10)
\Line(-60,0)(-40,0) 
\Line(-40,0)(-20,0)       
\DashLine(-40,0)(-40,20){3} 
\Line(-20,1.5)(0,1.5) \Line(-20,-1.5)(0,-1.5) 
\Line(0,1.5)(20,1.5)  \Line(0,-1.5)(20,-1.5)  
\Line(20,0)(40,0) 
\DashCArc(0,0)(20,0,180){4} 
\Vertex(-40,0){3} \Vertex(-20,0){3} \Vertex(20,0){3} 
\SetWidth{1}   \BBoxc(0,0)(5,5) 
\end{picture}
   \hspace*{\fill} 
\caption{\label{p-wave,loop}%
One-loop diagrams contributing to P-wave hyperon non-leptonic 
decay amplitudes, with weak vertices from  ${\cal L}^{\rm w}$ 
in  Eq.~(\ref{loweak}).}
\end{figure}
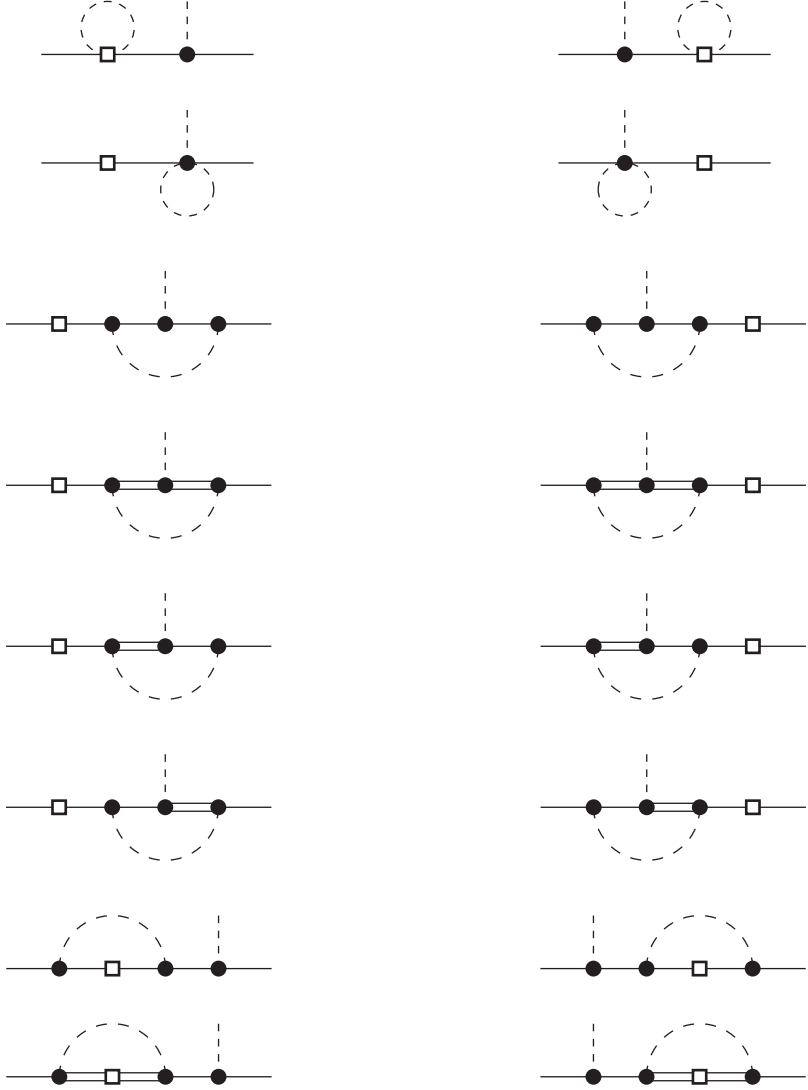             

We present our results for the ${\cal O}(m_s^{}\ln{m_s^{}})$ terms 
separating the contributions from different types of diagrams. 
From one-loop diagrams involving only octet baryons, shown in 
Figures~\ref{s-wave,loop}~and~\ref{p-wave,loop}, we obtain  
\vspace{1ex}           
\begin{eqnarray}         
\begin{array}{rlrl}   \displaystyle  
\beta^{(\rm S)}_{\Sigma^+ n}    &\!\!=\;   
\Bigl( \ratio{23}{8} -    
      \ratio{5}{4} D^2 - 3 D F + \ratio{9}{4} F^2 \Bigr)   
\beta_{27}^{}    \;, 
&   \hspace{3em}   
\beta^{(\rm S)}_{\Lambda p}     &\!\!=\; 0   \;,
\vspace{3ex} \\   \displaystyle    
\beta^{(\rm S)}_{\Sigma^- n}    &\!\!=\;   
\Bigl( -\ratio{23}{12}   
      + \ratio{5}{6} D^2 + 2 D F - \ratio{3}{2} F^2 \Bigr) 
\beta_{27}^{}    \;, &  
\beta^{(\rm S)}_{\Xi^-\Lambda}  &\!\!=\; 0   \;,  
\end{array}    
\label{octetsl}  
\end{eqnarray}
\begin{eqnarray}         
\begin{array}{rl}   \displaystyle  
\beta^{(\rm P)}_{\Sigma^+ n}  &\!\!=\;   \displaystyle   
\Bigl( \ratio{29}{24} D + \ratio{29}{8} F 
      -  \ratio{19}{36} D^3 - \ratio{29}{12} D^2 F 
      - \ratio{31}{12} D F^2 + \ratio{15}{4} F^3 \Bigr) 
{\beta_{27}^{}\over m_\Sigma^{}-m_N^{}}    \;,
\vspace{3ex} \\   \displaystyle    
\beta^{(\rm P)}_{\Sigma^- n}  &\!\!=\;   \displaystyle   
\Bigl( -\ratio{29}{12} F  
      + \ratio{8}{9} D^2 F + 2 D F^2 - 2 F^3 \Bigr) 
{\beta_{27}^{}\over m_\Sigma^{}-m_N^{}}       \;,
\vspace{3ex} \\   \displaystyle  
\beta^{(\rm P)}_{\Lambda p}  &\!\!=\;   \displaystyle  
\ratio{1}{\sqrt{6}} 
\Bigl( -\ratio{29}{12} D 
      + \ratio{16}{9} D^3 + 2 D^2 F - 2 D F^2 \Bigr)   
{\beta_{27}^{}\over m_\Sigma^{}-m_N^{}}       \;,
\vspace{3ex} \\   \displaystyle  
\beta^{(\rm P)}_{\Xi^-\Lambda}  &\!\!=\;   \displaystyle  
\ratio{1}{\sqrt{6}} 
\Bigl( \ratio{29}{12} D  
      - \ratio{16}{9} D^3 + 2 D^2 F + 2 D F^2 \Bigr)  
{\beta_{27}^{}\over m_\Xi-m_\Sigma}    \;.  
\end{array}   \label{octetpl}        
\end{eqnarray}
\hfill \\           
From one-loop diagrams involving decuplet baryons (also shown in 
Figures~\ref{s-wave,loop}~and~\ref{p-wave,loop}), we find  
\vspace{1ex}           
\begin{eqnarray}         
\begin{array}{rlrl}   \displaystyle     
\beta^{\prime(\rm S)}_{\Sigma^+ n}    &\!\!=\;   
-\ratio{1}{3} \,{\cal C}^2 \delta_{27}^{}  \;, & \hspace{3em}   
\beta^{\prime(\rm S)}_{\Lambda p}     &\!\!=\;  0   \;,
\vspace{3ex} \\   \displaystyle  
\beta^{\prime(\rm S)}_{\Sigma^- n}    &\!\!=\;   
\ratio{2}{9} \,{\cal C}^2 \delta_{27}^{}   \;, & 
\beta^{\prime(\rm S)}_{\Xi^-\Lambda}  &\!\!=\; 0   \;,    
\end{array}          
\end{eqnarray}
\begin{eqnarray}         
\begin{array}{rl}   \displaystyle   
\beta^{\prime(\rm P)}_{\Sigma^+ n}  &\!\!=\;   \displaystyle   
\Bigl( \ratio{85}{162} {\cal H}    
      \,-\, \ratio{35}{27} D + \ratio{1}{9} F \Bigr) 
{\cal C}^2\, {\beta_{27}^{}\over m_\Sigma^{}-m_N^{}}    
\,-\,   
\Bigl( \ratio{1}{9} D + \ratio{1}{3} F \Bigr) {\cal C}^2\, 
{\delta_{27}^{}\over m_\Sigma^{}-m_N^{}}       \;,
\vspace{3ex} \\   \displaystyle   
\beta^{\prime(\rm P)}_{\Sigma^- n}  &\!\!=\;   \displaystyle    
\Bigl( -\ratio{25}{54} {\cal H}    
      \,+\,  \ratio{26}{27} D - \ratio{2}{9} F \Bigr) 
{\cal C}^2\, {\beta_{27}^{}\over m_\Sigma^{}-m_N^{}}       
\,+\,    
\ratio{2}{9} F {\cal C}^2\, {\delta_{27}^{}\over m_\Sigma^{}-m_N^{}}   \;,
\vspace{3ex} \\   \displaystyle    
\beta^{\prime(\rm P)}_{\Lambda p}  &\!\!=\;   \displaystyle   
\ratio{1}{\sqrt{6}} 
\Bigl( -\ratio{5}{54} {\cal H} 
      + \ratio{4}{3} D + \ratio{4}{3} F \Bigr) 
{\cal C}^2\, {\beta_{27}^{}\over m_\Sigma^{}-m_N^{}}       
\,+\,    
\ratio{2}{9\sqrt{6}} D {\cal C}^2\, 
{\delta_{27}^{}\over m_\Sigma^{}-m_N^{}}   \;,   
\vspace{3ex} \\   \displaystyle     
\beta^{\prime(\rm P)}_{\Xi^-\Lambda}  &\!\!=\;   \displaystyle   
\ratio{1}{\sqrt{6}} 
\Bigl( \ratio{5}{54} {\cal H} 
      - \ratio{4}{3} D - \ratio{4}{3} F \Bigr)   
{\cal C}^2\, {\beta_{27}^{}\over m_\Xi-m_\Sigma}    
\,-\,    
\ratio{1}{\sqrt{6}} D {\cal C}^2\, 
{\delta_{27}^{}\over m_\Xi^{}-m_\Sigma^{}}   \;.  
\end{array}   
\end{eqnarray}

The contributions from the wave-function renormalization of 
the pion and the octet baryons and from the renormalization of 
the pion-decay constant are collected into        
\begin{eqnarray}         
\bar{\lambda}_{B_{}^{}B_{}'\pi}^{}  \;=\;  
\ratio{1}{2} \!\left( \bar{\lambda}_{B_{}^{}}^{} 
                   + \bar{\lambda}_{B_{}'}^{} 
                   + \lambda_\pi^{} \right) - \lambda_f^{} \;,       
\end{eqnarray}    
where  $\;\bar{\lambda}_{B}^{}=\lambda_{B}^{}+\lambda_{B}',\;$  
$\lambda_\pi^{}$  and  $\lambda_f^{}$  are defined by   
\begin{eqnarray}        
Z_B^{}  \;=\;  
1 + \bar{\lambda}_B^{}\, {m_K^2\over 16\pi^2 f_{\!\pi}^2}\, 
   \ln{m_K^2\over\mu^2}  
\;, \hspace{2em}   
Z_\pi^{}  \;=\;  
1 + \lambda_\pi^{}\, {m_K^2\over 16\pi^2 f_{\!\pi}^2}\, 
   \ln{m_K^2\over\mu^2}   \;,     
\end{eqnarray}    
and
\begin{eqnarray}        
{f_{\!\pi}^{}\over f}  \;=\;  
1 + \lambda_f^{}\, {m_K^2\over 16\pi^2 f_{\!\pi}^2}\, 
   \ln{m_K^2\over\mu^2}   \;, 
\end{eqnarray}    
with   
\begin{eqnarray}         
\begin{array}{rlrl}   \displaystyle  
\lambda_{N}^{}  &\!\!=\;   
\ratio{17}{6} D^2 - 5 D F + \ratio{15}{2} F^2   \;, & \hspace{3em}   
\lambda_{N}'    &\!\!=\;   \ratio{1}{2} \,{\cal C}^2   \;, 
\vspace{3ex} \\   \displaystyle  
\lambda_{\Lambda}^{}  &\!\!=\;  \ratio{7}{3} D^2 + 9 F^2 
\;, & \hspace{3em} 
\lambda_{\Lambda}'    &\!\!=\;  {\cal C}^2   \;,  
\vspace{3ex} \\   \displaystyle  
\lambda_{\Sigma}^{}   &\!\!=\;  \ratio{13}{3} D^2 + 3 F^2 
\;, & \hspace{3em}   
\lambda_{\Sigma}'     &\!\!=\;  \ratio{7}{3} \,{\cal C}^2   \;,  
\vspace{3ex} \\   \displaystyle  
\lambda_{\Xi}^{}  &\!\!=\;   
\ratio{17}{6} D^2 + 5 D F + \ratio{15}{2} F^2   \;, \hspace{3em} &  
\lambda_{\Xi}'    &\!\!=\;  \ratio{13}{6} \,{\cal C}^2   \;,  
\\     
\end{array}   
\end{eqnarray}   
\begin{eqnarray}         
\lambda_{\pi}^{}  \;=\;  -\ratio{1}{3}   
\;, \hspace{3em}   
\lambda_f^{}  \;=\;  -\ratio{1}{2}   \;.  
\end{eqnarray}    

Finally, for the P-waves, we must also include one-loop 
corrections to the propagator that appears in tree-level 
pole diagrams.  
These corrections have been partially addressed for 
the  $\;|\Delta\bfm{I}|=1/2\;$  amplitudes by  
Springer~\cite{springer}.  
We find that it is sufficient to take into account 
the one-loop renormalization of the baryon masses to correctly 
include these non-analytic corrections to the hyperon decay amplitudes. 
This is true for both the usual terms of  
${\cal O}(m_s^{}\ln{m_s^{}})$, as well as terms of   
${\cal O}(\pi m_s^{1/2})$. 
The latter correspond to the baryon-mass corrections that are  
proportional to $\pi m_K^3$ in the calculation of Ref.~\cite{jenmass}. 
We find  
\begin{eqnarray}   
\gamma_{\Sigma^+ n}^{}  \;=\;  \gamma_{\Sigma^- n}^{}  \;=\;   
\gamma_{\Lambda p}^{}  \;=\;  
{\mu_{\Sigma N}^{}\over m_\Sigma^{} - m_N^{}}   
\;, \hspace{3em}  
\gamma_{\Xi^-\Lambda}^{}  \;=\;  
{\mu_{\Xi\Sigma}^{}\over m_\Xi^{} - m_\Sigma^{}}   \;,     
\label{startmassr}
\end{eqnarray}      
where        
\begin{eqnarray}   \label{muxy}
\mu_{XY}^{}  &\!\!\!=&\!\!\!      
- \Bigl( \bar{\beta}_{X}^{}-\bar{\beta}_{Y}^{} \Bigr)   
 {m_K^3\over 16\pi f_{\!\pi}^2}   
\nonumber \\ && \!\!\! \!\!\!   
+\; 
\Bigl[ \Bigl( \bar{\gamma}_{X}^{} - \bar{\gamma}_{Y}^{} 
             - \bar{\lambda}_{X}^{}\alpha_{X}^{} 
             + \bar{\lambda}_{Y}^{}\alpha_{Y}^{} \Bigr) m_s^{}
      + \Bigl( \lambda_{X}' - \lambda_{Y}' \Bigr) \Delta m \Bigr]  
{m_K^2\over 16\pi^2 f_{\!\pi}^2}\, \ln{m_K^2\over\mu^2}   \;, 
\label{mkcubed} 
\end{eqnarray}      
and   
\begin{eqnarray}         
\alpha_{N}^{}  \;=\;  -2 \left( b_D^{}-b_F^{} \right) - 2\sigma   
\;, \hspace{2em} 
\alpha_{\Sigma}^{}  \;=\;  -2\sigma   
\;, \hspace{2em} &  
\alpha_{\Xi}^{}  \;=\;  -2 \left( b_D^{}+b_F^{} \right) - 2\sigma   \;,
\end{eqnarray}   
\begin{eqnarray}         
\begin{array}{rl}   \displaystyle  
\bar{\beta}_{N}^{}  &\!\!=\;   
\ratio{5}{3} D^2 - 2 D F + 3 F^2 
+ \ratio{4}{9\sqrt{3}} \!\left( D^2 - 6 D F + 9 F^2 \right)   
\;+\;   
\ratio{1}{3} \,{\cal C}^2   \;,  
\vspace{3ex} \\   \displaystyle  
\bar{\beta}_{\Sigma}^{}  &\!\!=\;   
2 D^2 + 2 F^2 + \ratio{16}{9\sqrt{3}} D^2 
\;+\;
\left( \ratio{10}{9} + \ratio{8}{9\sqrt{3}} \right) {\cal C}^2   \;, 
\vspace{3ex} \\   \displaystyle  
\bar{\beta}_{\Xi}^{}  &\!\!=\;   
\ratio{5}{3} D^2 + 2 D F + 3 F^2 
+ \ratio{4}{9\sqrt{3}} \!\left( D^2 + 6 D F + 9 F^2 \right) 
\;+\;  
\left( 1 + \ratio{8}{9\sqrt{3}} \right) {\cal C}^2   \;,     
\end{array}   
\end{eqnarray}   
\begin{eqnarray}         
\begin{array}{rl}   \displaystyle  
\bar{\gamma}_{N}^{}  &\!\!=\;   
\ratio{43}{9}\, b_D^{} - \ratio{25}{9}\, b_F^{} 
- b_D^{} \left( \ratio{4}{3} D^2 + 12 F^2 \right)  
+ b_F^{} \left( \ratio{2}{3} D^2 - 4 D F + 6 F^2 \right)  
+ \ratio{52}{9}\, \sigma - 2 \sigma \lambda_N^{}     
\vspace{1ex} \\   \displaystyle  
& \;+\; 
\ratio{1}{3}\, c \,{\cal C}^2 - 2 \tilde{\sigma} \lambda_N'   \;,   
\vspace{3ex} \\   \displaystyle  
\bar{\gamma}_{\Sigma}^{}  &\!\!=\;   
2 b_D^{} 
- b_D^{} \left( 6 D^2 + 6 F^2 \right)  
- b_F^{} \left( 12 D F \right)  
+ \ratio{52}{9}\, \sigma - 2 \sigma \lambda_\Sigma^{}   
\;+\;    
\ratio{8}{9}\, c \,{\cal C}^2 - 2 \tilde{\sigma} \lambda_\Sigma'   \;,  
\vspace{3ex} \\   \displaystyle  
\bar{\gamma}_{\Xi}^{}  &\!\!=\;   
\ratio{43}{9}\, b_D^{} + \ratio{25}{9}\, b_F^{} 
- b_D^{} \left( \ratio{4}{3} D^2 + 12 F^2 \right)  
- b_F^{} \left( \ratio{2}{3} D^2 + 4 D F + 6 F^2 \right)  
+ \ratio{52}{9}\, \sigma - 2 \sigma \lambda_\Xi^{}   
\vspace{1ex} \\   \displaystyle  
& \;+\;  
\ratio{29}{9}\, c \,{\cal C}^2 - 2 \tilde{\sigma} \lambda_\Xi'   \;.     
\end{array}
\label{endmassr}   
\end{eqnarray}

The parameters that appear in the strong Lagrangian, 
Eq.~(\ref{L1strong}),  can be measured in semi-leptonic hyperon decays.   
For our numerical estimates we will employ the values
\begin{eqnarray}         
D  \;=\;  0.61 \pm 0.04  \;, \hspace{2em}   
F  \;=\;  0.40 \pm 0.03  \;, \hspace{2em}   
{\cal C}  \;=\;  1.6   \;, \hspace{2em}  
{\cal H}  \;=\;  -1.9 \pm 0.7  \;,    
\label{semifit}     
\end{eqnarray}    
which were obtained from a three-parameter fit to semi-leptonic 
hyperon decays including non-analytic corrections from octet- and 
decuplet-baryon loops~\cite{manjenco}.

The values of the parameters that appear in  Eq.~(\ref{Lmstrong}) 
may be obtained by fitting the tree-level expressions for the baryon  
masses  [including terms of up to  ${\cal O}(m_s^{})$]  to  
the physical masses. 
At this order, the baryon-octet masses are not independent, 
and instead they satisfy the Gell-Mann-Okubo relation.       
Similarly, the baryon-decuplet masses satisfy Gell-Mann's   
equal-spacing rule.   
All this implies that it is possible to extract these parameters 
in more than one way from the physical masses.  
The different parameter sets obtained are equivalent to   
${\cal O}(m_s^{})$.   
We choose the following, 
\vspace{1ex}           
\begin{eqnarray}   \label{newmasspar}     
\begin{array}{rl}   \displaystyle  
b_D^{} m_s^{}  &\!\!=\;  
\ratio{3}{8} \left( m_\Sigma^{}-m_\Lambda^{} \right)  
\;\approx\;  0.0290\,{\rm GeV}   \;,  
\vspace{3ex} \\   \displaystyle   
b_F^{} m_s^{}  &\!\!=\;  \ratio{1}{4} \left( m_N^{}-m_\Xi^{} \right)  
\;\approx\;  -0.0948\,{\rm GeV}   \;,  
\vspace{3ex} \\   \displaystyle   
c m_s^{}  &\!\!=\;  \ratio{1}{2} \left( m_\Omega^{}-m_\Delta^{} \right)   
\;\approx\;  0.220\,{\rm GeV}   \;,  
\vspace{3ex} \\   \displaystyle   
\Delta m - 2 \left( \tilde{\sigma}-\sigma \right) m_s^{}   &\!\!=\;  
m_\Delta^{}-m_\Sigma^{}   
\;\approx\;  0.0389\,{\rm GeV}   \;,     
\end{array}   
\end{eqnarray}    
\hfill \\          
where the fourth parameter is the only combination of 
$\Delta m$,  $\sigma m_s^{}$  and  $\tilde{\sigma} m_s^{}$   
which occurs in  Eq.~(\ref{muxy}).

\section{Beyond the Leading Non-Analytic Corrections}

In the previous section we have calculated the terms of  
${\cal O}(m_s^{}\ln{m_s^{}})$ that occur in the  
$\;|\Delta\bfm{I}|=3/2\;$  hyperon non-leptonic decay amplitudes.  
To this order we have found that the S-wave amplitudes for 
$\Lambda$ and $\Xi$ decays vanish, whereas the experimentally measured 
amplitudes do not. In this section we discuss
two examples of terms that contribute to these amplitudes and use 
them as an estimate of their size in $\chi$PT. 
A prediction for these amplitudes is not possible at present due 
to the unknown constants that occur at the order 
at which we first find a non-zero result,~${\cal O}(m_s^{})$.

An example of a non-vanishing contribution at  ${\cal O}(m_s^{})$ 
is that from a counterterm already discussed in  Ref.~\cite{heval}. 
In the heavy-baryon formalism, it corresponds to
\begin{eqnarray}   
{\cal L}_1^{\rm w}  \;=\;     
{\tilde{\beta}_{27}^{}\over m_N^{}}\, 
T_{ij,kl}^{} \left[ 
\left( \xi\bar{B}_v^{}\right)_{\!kn}^{} 
\left( B_v^{} \xid \right) _{\!ni}^{} 
- \left( \bar{B}_v^{} \xid \right) _{\!ni}^{}  
 \left( \xi B_v^{} \right) _{\!kn}^{}  
\right] 
\left( \ri v^\mu\partial_\mu^{}\Sigma\, 
      \Sigma^\dagger \right) _{\!lj}^{}  
\;+\;  {\rm h.c.}   \;,      
\label{higheror}
\end{eqnarray}      
and gives rise to the contributions 
\vspace{1ex}          
\begin{eqnarray}        
\begin{array}{rlrl}   \displaystyle  
\alpha^{(\rm S)}_{\Sigma^+ n}  &\!\!=\;  0       
\;, &   
\alpha^{(\rm S)}_{\Lambda p}   &\!\!=\;   \displaystyle  
-\sqrt{\ratio{3}{2}}\, \tilde{\beta}_{27}^{}\, 
{m_\Lambda^{}-m_N^{} \over m_N^{}}   \;,   
\vspace{3ex} \\   \displaystyle   
\alpha^{(\rm S)}_{\Sigma^- n}    &\!\!=\;   \displaystyle  
-\tilde{\beta}_{27}^{}\, {m_\Sigma^{}-m_N^{} \over m_N^{}}   
\;, & \hspace{2em} 
\alpha^{(\rm S)}_{\Xi^-\Lambda}  &\!\!=\;   \displaystyle  
\sqrt{\ratio{3}{2}}\, \tilde{\beta}_{27}^{}\, 
{m_\Xi^{}-m_\Lambda^{} \over m_N^{}}   \;.           
\end{array}   
\end{eqnarray}    
\hfill \\        
Eq.~(\ref{higheror}) has been normalized so that  
$\tilde{\beta}_{27}^{}$  is naturally of the same order as  
$\beta_{27}^{}$.\footnote{We have used $m_N^{}$ as a normalization  
scale, but this is not meant to imply that this is a heavy-baryon 
mass correction. 
It could just as well be the chiral-symmetry breaking scale, 
$\Lambda_{\chi SB}^{}$.}
As remarked in  Ref.~\cite{heval}, 
this operator reproduces the current-current form of  
vacuum saturation for the S-waves if one takes  
$\;\tilde{\beta}_{27}^{}\approx 
   -0.32 \,\sqrt{2}\, f_{\!\pi}^{} G_{\rm F}^{} m_{\pi}^2.\;$   
We will find in the following section that a lowest-order fit to
S-wave  $\Sigma$  decays gives    
$\;\beta_{27}^{}\approx 
   -0.07\,\sqrt{2}\, f_{\!\pi}^{}G_{\rm F}^{} m_{\pi}^2,\;$  
and, therefore, the vacuum-saturation value for  
$\tilde{\beta}_{27}^{}$ is five times larger than expected.   
There is no reason, however, to 
trust the vacuum-saturation approximation, and we prefer to use 
$\;\tilde{\beta}_{27}^{}\approx \beta_{27}^{}\;$  to estimate 
these terms. 
In the literature one finds that vacuum-saturation calculations 
often include a meson-pole diagram for the P-waves.   
In particular,  Ref.~\cite{pakvasa} claims that these pole 
diagrams are important to fit experiment.       
Within the framework of  $\chi$PT,  these contributions are  
suppressed with respect to the ones we calculate by factors of  
$\;m_\pi^2/m_K^2\;$ or $\;m_d^{}/m_s^{}.\;$     
We see that the  $\chi$PT  predictions are 
completely at odds with those of  vacuum saturation for 
hyperon decays.

Other non-vanishing contributions to the S-wave 
$\Lambda$ and $\Xi$ decays come from the one-loop diagrams 
in which the weak transition involves only mesons, as in 
Figure~\ref{g27,loop}.   
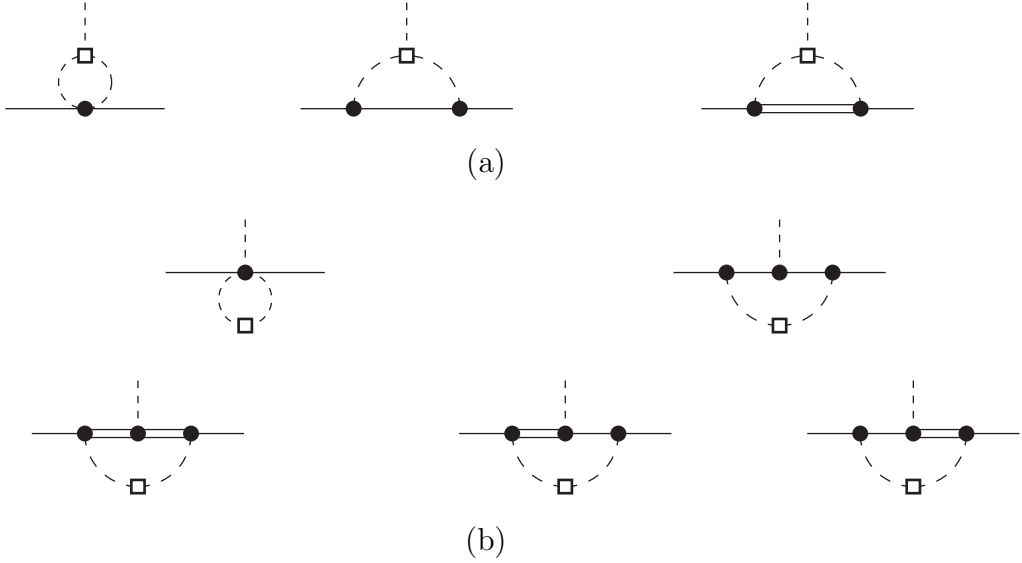
\begin{figure}[bt]         
   \hspace*{\fill} 
\begin{picture}(60,70)(-30,-10)
\Line(-30,0)(30,0) \Vertex(0,0){3}         
\DashCArc(0,10)(10,0,360){3}        
\DashLine(0,20)(0,40){3}    
\SetWidth{1}   \BBoxc(0,20)(5,5)        
\end{picture}
   \hspace*{\fill} 
\begin{picture}(80,70)(-20,-10)
\Line(-40,0)(40,0) 
\DashCArc(0,0)(20,0,180){4} 
\Vertex(-20,0){3} \Vertex(20,0){3} 
\DashLine(0,20)(0,40){3}    
\SetWidth{1}   \BBoxc(0,20)(5,5) 
\end{picture}
   \hspace*{\fill} 
\begin{picture}(80,70)(-20,-10)
\Line(-40,0)(-20,0) 
\Line(-20,1.5)(20,1.5) \Line(-20,-1.5)(20,-1.5) 
\Line(20,0)(40,0) 
\DashCArc(0,0)(20,0,180){4} 
\Vertex(-20,0){3} \Vertex(20,0){3} 
\DashLine(0,20)(0,40){3}    
\SetWidth{1}   \BBoxc(0,20)(5,5) 
\end{picture}
   \hspace*{\fill} 
\\   
   \hspace*{\fill} 
\begin{picture}(20,20)(-10,-10)
\Text(0,0)[c]{(a)}   
\end{picture}
   \hspace*{\fill} 
\\   
   \hspace*{\fill} 
\begin{picture}(120,60)(-60,-30)    
\Line(-30,0)(0,0) \Vertex(0,0){3} 
\DashLine(0,0)(0,20){3} \DashCArc(0,-10)(10,90,270){3}        
\DashCArc(0,-10)(10,-90,90){3}        
\Line(0,0)(30,0)   
\SetWidth{1}   \BBoxc(0,-20)(5,5) 
\end{picture}
   \hspace*{\fill} 
\begin{picture}(80,60)(-40,-30)    
\Line(-40,0)(-20,0)   
\Line(-20,0)(0,0) \Line(0,0)(20,0) \Line(20,0)(40,0)   
\Vertex(-20,0){3} \Vertex(0,0){3} \Vertex(20,0){3}   
\DashCArc(0,0)(20,180,270){4} \DashCArc(0,0)(20,270,0){4}   
\DashLine(0,0)(0,20){3}    
\SetWidth{1}   \BBoxc(0,-20)(5,5)   
\end{picture}   
   \hspace*{\fill}   
\\  
   \hspace*{\fill} 
\begin{picture}(140,60)(-70,-30)   
\Line(-40,0)(-20,0) 
\Line(-20,1.5)(0,1.5) \Line(-20,-1.5)(0,-1.5) 
\Line(0,1.5)(20,1.5)  \Line(0,-1.5)(20,-1.5)  
\Line(20,0)(40,0) 
\Vertex(-20,0){3} \Vertex(0,0){3} \Vertex(20,0){3} 
\DashCArc(0,0)(20,180,270){4} \DashCArc(0,0)(20,270,0){4} 
\DashLine(0,0)(0,20){3}   
\SetWidth{1}   \BBoxc(0,-20)(5,5) 
\end{picture}
   \hspace*{\fill} 
\begin{picture}(80,60)(-40,-30)
\Line(-40,0)(-20,0) 
\Line(-20,1.5)(0,1.5) \Line(-20,-1.5)(0,-1.5) 
\Line(0,0)(20,0)  \Line(20,0)(40,0) 
\Vertex(-20,0){3} \Vertex(0,0){3} \Vertex(20,0){3} 
\DashCArc(0,0)(20,180,270){4} \DashCArc(0,0)(20,270,0){4} 
\DashLine(0,0)(0,20){3} 
\SetWidth{1}   \BBoxc(0,-20)(5,5) 
\end{picture}
   \hspace*{\fill} 
\begin{picture}(80,60)(-40,-30)
\Line(-40,0)(-20,0) \Line(-20,0)(0,0)  
\Line(0,1.5)(20,1.5) \Line(0,-1.5)(20,-1.5) 
\Line(20,0)(40,0) 
\Vertex(-20,0){3} \Vertex(0,0){3} \Vertex(20,0){3} 
\DashCArc(0,0)(20,180,270){4} \DashCArc(0,0)(20,270,0){4} 
\DashLine(0,0)(0,20){3} 
\SetWidth{1}   \BBoxc(0,-20)(5,5) 
\end{picture}
   \hspace*{\fill} 
\\   
   \hspace*{\fill} 
\begin{picture}(20,20)(-10,-10)
\Text(0,0)[c]{(b)}   
\end{picture}
   \hspace*{\fill} 
\caption{\label{g27,loop}%
One-loop diagrams contributing to (a) S-wave and (b) P-wave 
hyperon non-leptonic decay amplitudes, with weak vertices from   
${\cal L}^{\rm w}_\phi$  in  Eq.~(\ref{mesonwe}).}  
\end{figure}             
These diagrams give calculable 
contributions that are formally of  ${\cal O}(m_s^2\ln{m_s^{}})$ 
and, therefore, should be smaller than the contributions of  
${\cal O}(m_s^{}\ln{m_s^{}})$. The corresponding diagrams for 
$\;|\Delta\bfm{I}|=1/2\;$ transitions were included in the 
calculation of Jenkins \cite{jenkins}, who argued that the 
large value of $g_8^{}$ in Eq.~(\ref{mesonweoc})  
(corresponding to her $h_\pi^{}$) could compensate their 
being of higher order.  
At present we do not have a detailed understanding of the 
$\;|\Delta\bfm{I}|=1/2\;$ enhancement in kaon decays so we 
cannot really rule out Jenkins' argument.   
However, it is also possible that whatever is responsible for 
the enhancement of $\;|\Delta\bfm{I}|=1/2\;$ kaon decays will also   
enhance  $\;|\Delta\bfm{I}|=1/2\;$  baryon decays in a similar way, 
invalidating Jenkins' argument.    
In other words, the coefficients $h_D^{}$ and $h_F^{}$ of 
Eq.~(\ref{loweakoc}) are also enhanced with respect to 
the expectation of naive dimensional analysis.  
In either case, the numerical results of Jenkins confirm that 
the non-analytic contributions from the diagrams analogous to   
those in Figure~\ref{g27,loop} are important 
for the  $\;|\Delta\bfm{I}|=1/2\;$  amplitudes.\footnote{We  
disagree with the expressions presented in Ref.~\cite{jenkins} for 
these terms  (those proportional to $h_\pi^{}$ in  
Ref.~\cite{jenkins}).  
This disagreement, however, does not affect our present discussion. 
We will present our results for  $\;|\Delta\bfm{I}|=1/2\;$  
transitions elsewhere.} 
Notice that these terms do not involve unknown parameters so they 
can be quantified.

Returning to the $\;|\Delta\bfm{I}|=3/2\;$ amplitudes, we want 
to calculate from the diagrams in Figure~\ref{g27,loop}  
the terms proportional to the constant $g_{27}^{}$ of   
Eq.~(\ref{mesonwe}).  
In this case, the dimensionless constant  
$\;g_{27}^{}\approx 0.16,\,$  unlike $g_8^{}$,  is suppressed 
with respect to the expectation from naive dimensional analysis.  
These terms, of  ${\cal O}(m_s^2\ln{m_s^{}})$, would 
be further suppressed by the small value of $g_{27}^{}$ and 
completely negligible if Jenkins' argument to include the analogous 
terms is correct. However, if the $\;|\Delta\bfm{I}|=1/2\;$ enhancement 
is universal, the large value of $g_8^{}$ is not responsible for the 
importance of these terms in the calculation of Jenkins. 
In this case,  we expect that these terms could be equally important  
for the  $\;|\Delta\bfm{I}|=3/2\;$ amplitudes.   
A posteriori, we find that these terms are indeed as important as  
those of  ${\cal O}(m_s^{}\ln{m_s^{}})$ as it happened in  
the calculation of the  $\;|\Delta\bfm{I}|=1/2\;$  amplitudes. 
We will return to this discussion in Section~6. 
Here we present the analytical expression for these terms.

\newpage              

The contributions to the amplitudes of
Eqs.~(\ref{defswave}) and~(\ref{defpwave}) 
from diagrams involving only octet baryons are   
\vspace{1ex}        
\begin{eqnarray}         
\begin{array}{rl}   \displaystyle  
\beta^{(\rm S)}_{\Sigma^+ n}  &\!\!=\;         
-\Bigl[ \left( 3 D^2 + 9 D F \right) (m_\Lambda^{}-m_N^{})  
      + \left( 7 D^2 - 9 D F \right) (m_\Sigma^{}-m_N^{}) \Bigr] 
\gamma_{27}^{}   \;,        
\vspace{3ex} \\   \displaystyle   
\beta^{(\rm S)}_{\Sigma^- n}  &\!\!=\;              
\Bigl[ \left( -3 - D^2 + 12 D F - 9 F^2 \right) (m_\Sigma^{}-m_N^{})  
      + \left( 2 D^2 + 6 D F \right) (m_\Lambda^{}-m_N^{}) \Bigr]    
\gamma_{27}^{}   \;,  
\vspace{3ex} \\   \displaystyle   
\beta^{(\rm S)}_{\Lambda p}  &\!\!=\;              
-\ratio{1}{\sqrt{6}} \!\left( 9 + 13 D^2 + 18 D F + 27 F^2 \right) 
\bigl( m_\Lambda^{}-m_N^{} \bigr) \, \gamma_{27}^{}   \;,    
\vspace{3ex} \\   \displaystyle   
\beta^{(\rm S)}_{\Xi^-\Lambda}  &\!\!=\;              
\ratio{1}{\sqrt{6}} \!\left( 9 + 13 D^2 - 18 D F + 27 F^2 \right) 
\bigl( m_\Xi^{}-m_\Lambda^{} \bigr) \, \gamma_{27}^{}   \;,    
\end{array}         
\end{eqnarray}         
\begin{eqnarray}         
\begin{array}{rl}   \displaystyle  
\vspace{-1ex} \\   
\beta^{(\rm P)}_{\Sigma^+ n}  &\!\!=\;              
\Bigl( \ratio{1}{3} D^3 + \ratio{7}{3} D^2 F + D F^2 - F^3 \Bigr)   
\gamma_{27}^{}   \;,      
\vspace{3ex} \\   \displaystyle   
\beta^{(\rm P)}_{\Sigma^- n}  &\!\!=\;         
\Bigl( \ratio{1}{3} D - \ratio{1}{3} F 
      - D^3 - \ratio{5}{3} D^2 F - 3 D F^2 + 3 F^3 \Bigr)   
\gamma_{27}^{}   \;,      
\vspace{3ex} \\   \displaystyle   
\beta^{(\rm P)}_{\Lambda p}  &\!\!=\;              
\ratio{1}{\sqrt{6}} \Bigl( 
-\ratio{1}{3} D - F  
+ \ratio{5}{3} D^3 + 9 D^2 F + 5 D F^2 + 3 F^3 
\Bigr) \gamma_{27}^{}   \;,      
\vspace{3ex} \\   \displaystyle   
\beta^{(\rm P)}_{\Xi^-\Lambda}  &\!\!=\;  
\ratio{1}{\sqrt{6}} \Bigl(   
-\ratio{1}{3} D + F  
+ \ratio{5}{3} D^3 - 9 D^2 F + 5 D F^2 - 3 F^3 
\Bigr) \gamma_{27}^{}   \;.        
\end{array}
\end{eqnarray}
\hfill \\        
From diagrams involving decuplet baryons, we find
\vspace{1ex}        
\begin{eqnarray}         
\begin{array}{rl}   \displaystyle  
\beta^{\prime(\rm S)}_{\Sigma^+ n}  &\!\!=\;          
{\cal C}^2\, 
\Bigl[ \ratio{3}{2} (m_\Sigma^{}-m_N^{}) - 2(m_\Delta^{}-m_N^{}) 
      + m_{\Sigma^*}^{}-m_N^{} \Bigr] \, \gamma_{27}^{}   \;,  
\vspace{3ex} \\   \displaystyle   
\beta^{\prime(\rm S)}_{\Sigma^- n}  &\!\!=\;              
{\cal C}^2\, 
\Bigl[ -\ratio{23}{6} (m_\Sigma^{}-m_N^{})  
      + \ratio{4}{3} (m_\Delta^{}-m_N^{}) 
      - \ratio{2}{3} (m_{\Sigma^*}^{}-m_N^{}) \Bigr] \, \gamma_{27}^{}   \;,  
\vspace{3ex} \\   \displaystyle   
\beta^{\prime(\rm S)}_{\Lambda p}  &\!\!=\;         
\ratio{5}{2\sqrt{6}} \,{\cal C}^2\, 
\bigl( m_\Lambda^{}-m_N^{} \bigr) \, \gamma_{27}^{}   \;,    
\hspace{3em} 
\beta^{\prime(\rm S)}_{\Xi^-\Lambda}  \;=\;  
\ratio{7}{2\sqrt{6}} \,{\cal C}^2\, 
\bigl( m_\Xi^{}-m_\Lambda^{} \bigr) \, \gamma_{27}^{}   \;,    
\end{array}
\end{eqnarray}
\begin{eqnarray}   
\begin{array}{rlrl}   \displaystyle  
\vspace{-1ex} \\   
\beta^{\prime(\rm P)}_{\Sigma^+ n}  &\!\!=\;   \displaystyle  
- \Bigl( \ratio{10}{81} {\cal H} 
      \,+\, \ratio{2}{9} D + \ratio{10}{9} F \Bigr) 
{\cal C}^2\, \gamma_{27}^{}   \;,   & \hspace{3em}        
\beta^{\prime(\rm P)}_{\Lambda p}   &\!\!=\;   \displaystyle      
-\ratio{1}{\sqrt{6}} \Bigl( 
\ratio{10}{27} {\cal H} + \ratio{8}{9} D+4 F 
\Bigr) {\cal C}^2\, \gamma_{27}^{}   \;,      
\vspace{3ex} \\   \displaystyle  
\beta^{\prime(\rm P)}_{\Sigma^- n}    &\!\!=\;   \displaystyle    
\Bigl( \ratio{10}{27} {\cal H} \,+\, \ratio{4}{3} D \Bigr) 
{\cal C}^2\, \gamma_{27}^{}   \;,   &     
\beta^{\prime(\rm P)}_{\Xi^-\Lambda}  &\!\!=\;   \displaystyle  
-\ratio{1}{\sqrt{6}} \Bigl( \ratio{8}{9} D - \ratio{4}{3} F \Bigr) 
{\cal C}^2\, \gamma_{27}^{}   \;.   
\end{array}     
\end{eqnarray}  
\hfill \\        
We have used the notation  
$\;\gamma_{27}^{}=G_{\rm F}^{} f_{\!\pi}^2 V_{ud}^{*} V_{us}^{}\,   
g_{27}^{}/ \sqrt{2} \approx  2.5 \times 10^{-9}$.

\section{Current Experimental Values}

From the measurement of the decay rate and the decay-distribution 
asymmetry parameter $\alpha$, it is possible to extract the value 
of the S- and P-wave amplitudes for each hyperon 
decay~\cite{dogoho}.       
Using the numbers from the 1998 Review of Particle 
Physics~\cite{pdb}, we find the results presented in 
Table~\ref{s,p,isosymmetric}.   
\begin{table}[b]      
\caption{\label{s,p,isosymmetric}%
Experimental values for S- and P-wave amplitudes.   
In extracting these numbers, final-state interactions have been 
ignored and isospin-symmetric masses used.}      
\centering   \small   
\vskip 1\baselineskip
\begin{tabular}{ccc}    
\hline \hline      
\raisebox{-1ex}{Decay mode $\hspace{1ex}$}   &    
\raisebox{-1ex}{$\hspace{1ex}s$}   &
\raisebox{-1ex}{$\hspace{1ex}p$}   
\vspace{1ex} \\ \hline      
\vspace{-2.5ex} & & \\  
$\begin{array}{rcl}   \displaystyle  
\Sigma^+  & \hspace{-.5em} \rightarrow & \hspace{-.5em}  n\pi^+ 
\\   
\Sigma^+  & \hspace{-.5em} \rightarrow & \hspace{-.5em}  p\pi^0   
\\      
\Sigma^-  & \hspace{-.5em} \rightarrow & \hspace{-.5em}  n\pi^-   
\\    
\Lambda   & \hspace{-.5em} \rightarrow & \hspace{-.5em}  p\pi^-    
\\        
\Lambda   & \hspace{-.5em} \rightarrow & \hspace{-.5em}  n\pi^0    
\\        
\Xi^-     & \hspace{-.5em} \rightarrow & \hspace{-.5em}  \Lambda\pi^-
\\  
\Xi^0     & \hspace{-.5em} \rightarrow & \hspace{-.5em}  \Lambda\pi^0
\end{array}$  
&  
$\begin{array}{rcl}   \displaystyle  
0.06   & \hspace{-.5em} \pm & \hspace{-.5em}  0.01   \\  
-1.48  & \hspace{-.5em} \pm & \hspace{-.5em}  0.05   \\  
1.95   & \hspace{-.5em} \pm & \hspace{-.5em}  0.01   \\  
1.46   & \hspace{-.5em} \pm & \hspace{-.5em}  0.01   \\  
-1.09  & \hspace{-.5em} \pm & \hspace{-.5em}  0.02   \\  
-2.06  & \hspace{-.5em} \pm & \hspace{-.5em}  0.01   \\ 
1.55   & \hspace{-.5em} \pm & \hspace{-.5em}  0.02   
\end{array}$  
&  
$\begin{array}{rcl}   \displaystyle   
1.85   & \hspace{-.5em} \pm & \hspace{-.5em}  0.01   \\  
1.21   & \hspace{-.5em} \pm & \hspace{-.5em}  0.06   \\  
-0.07  & \hspace{-.5em} \pm & \hspace{-.5em}  0.01   \\  
0.53   & \hspace{-.5em} \pm & \hspace{-.5em}  0.01   \\  
-0.40  & \hspace{-.5em} \pm & \hspace{-.5em}  0.03   \\  
0.50   & \hspace{-.5em} \pm & \hspace{-.5em}  0.02   \\ 
-0.33  & \hspace{-.5em} \pm & \hspace{-.5em}  0.02   
\end{array}$ 
\vspace{.3ex} \\  
\hline \hline 
\end{tabular}   
\vskip 1\baselineskip
\end{table}
Here  $s$  and  $p$  are related to  
${\cal A}^{\rm (S,P)}_{B_{}^{}B_{}'\pi}$  by   
\begin{eqnarray}   
s  \;=\;  {\cal A}^{\rm (S)}   \;, \hspace{2em}   
p  \;=\;  -|\bfm{k}| {\cal A}^{\rm(P)}   \;,      
\end{eqnarray}      
where  $\bfm{k}$  is the pion three-momentum in the rest frame of  
the decaying baryon.    
These numbers are very similar to those quoted 
by Jenkins~\cite{jenkins}  (whose conventions we follow), because 
there are no new experimental measurements. 
Our minor differences are due to the fact that we use masses in 
the isospin-symmetry limit instead of physical masses.  
From the amplitudes in  Table~\ref{s,p,isosymmetric}, we can 
extract the  $\;|\Delta\bfm{I}|=3/2\;$  components using 
the relations
\vspace{1ex}        
\begin{eqnarray}   
\begin{array}{c}   \displaystyle        
S_{3}^{(\Lambda)}  \;=\;  
\ratio{1}{\sqrt{3}} 
\Bigl( \sqrt{2}\, s_{\Lambda\rightarrow n\pi^0}^{}   
      + s_{\Lambda\rightarrow p\pi^-}^{} \Bigr)   \;,    
\vspace{3ex} \\   \displaystyle  
S_{3}^{(\Xi)}  \;=\;  
\ratio{2}{3} \Bigl( \sqrt{2}\, s_{\Xi^0\rightarrow\Lambda\pi^0}^{}  
                   + s_{\Xi^-\rightarrow\Lambda\pi^-}^{} \Bigr)   \;,  
\vspace{3ex} \\   \displaystyle  
S_{3}^{(\Sigma)}  \;=\;  
-\sqrt{\ratio{5}{18}} 
\Bigl( s_{\Sigma^+\rightarrow n\pi^+}^{}  
      - \sqrt{2}\, s_{\Sigma^+\rightarrow p\pi^0}^{}   
      - s_{\Sigma^-\rightarrow n\pi^-}^{} \Bigr)   \;.     
\end{array}   
\end{eqnarray}      
\hfill \\          
and corresponding ones for the P-wave amplitudes.   
Similarly, one obtains the  $\;|\Delta\bfm{I}|=1/2\;$  
components of the amplitudes (for  $\Lambda$  and  $\Xi$  decays)  
using the expressions  
\vspace{1ex}     
\begin{eqnarray}   
\begin{array}{c}   \displaystyle        
S_{1}^{(\Lambda)}  \;=\;  
\ratio{1}{\sqrt{3}} 
\Bigl( s_{\Lambda\rightarrow n\pi^0}^{}   
      - \sqrt{2}\, s_{\Lambda\rightarrow p\pi^-}^{} \Bigr)   \;,    
\vspace{3ex} \\   \displaystyle  
S_{1}^{(\Xi)}  \;=\;  
\ratio{\sqrt{2}}{3} 
\Bigl( s_{\Xi^0\rightarrow\Lambda\pi^0}^{}  
       - \sqrt{2}\, s_{\Xi^-\rightarrow\Lambda\pi^-}^{} \Bigr)   \;, 
\end{array}   
\end{eqnarray}      
\hfill \\        
and analogous ones for the P-waves. 
The  $\;|\Delta\bfm{I}|=1/2\;$  rule for hyperon decays can be seen 
in the ratios shown in  Table~\ref{ratio,isosymmetric}.
The experimental values for  $S_3^{}$  and  $P_3^{}$  
are listed in the column labeled ``Experiment'' in  
Table~\ref{results}.
\begin{table}[bt]      
\caption{\label{ratio,isosymmetric}%
Ratios of  $\;|\Delta\bfm{I}|=3/2\;$  to  $\;|\Delta\bfm{I}|=1/2\;$  
amplitudes, derived from Table \ref{s,p,isosymmetric}.   
Here  $\;S_-^{(\Sigma)}=s_{\Sigma^-\rightarrow n\pi^-}^{}\;$  
and  $\;P_+^{(\Sigma)}=p_{\Sigma^+\rightarrow n\pi^+}^{}.\;$}      
\centering   \small
\vskip 1\baselineskip   
\begin{tabular}{|c|r|} 
\hline \hline      
& \\ 
$S_3^{(\Lambda)}/S_{1_{}}^{(\Lambda)^{}}$   &   $0.026\pm 0.009$   
\\  
& \\ 
$P_3^{(\Lambda)}/P_1^{(\Lambda)}$   &   $0.031\pm 0.037$  
\\  
& \\ 
$S_3^{(\Xi)}/S_1^{(\Xi)}$           &   $0.042\pm 0.009$   
\\ 
& \\
$P_3^{(\Xi)}/P_1^{(\Xi)}$           &   $-0.045\pm 0.047$  
\\ 
& \\
$S_3^{(\Sigma)}/S_-^{(\Sigma)}$     &   $-0.055\pm 0.020$  
\\ 
& \\
$P_3^{(\Sigma)}/P_+^{(\Sigma)}$     &   $-0.059\pm 0.024$  
\\ 
& \\  
\hline \hline 
\end{tabular}   
\vskip 1\baselineskip
\end{table}

\section{Discussion}

In this section we present a numerical comparison of our results 
with experiment. We start with the tree-level, lowest-order in 
$\chi$PT, calculation.
At this order, there are four non-zero predictions that depend only  
on the parameter $\beta_{27}^{}$ (and on parameters from the 
strong sector that have been determined elsewhere). 
We can extract the value of $\beta_{27}^{}$  from each of 
these four amplitudes, and compare these values to test the 
consistency of the framework. We find, in units of  
$\;\sqrt{2}\, f_{\!\pi}^{}G_{\rm F}^{} m_{\pi}^2$,  
\begin{equation}
\beta_{27}^{}  \;=\;  
\left\{ \begin{array}{rlrl}   \displaystyle  
-0.068\,\pm  & \!\!\!\! 0.024  &  {\rm from}  & S_3^{(\Sigma)}  
\vspace{1ex} \\  
 0.12 \,\pm  & \!\!\!\! 0.15   &  {\rm from}  & P_3^{(\Lambda)} 
\vspace{1ex} \\  
 0.040\,\pm  & \!\!\!\! 0.042  &  {\rm from}  & P_3^{(\Xi)}     
\vspace{1ex} \\  
 0.23 \,\pm  & \!\!\!\! 0.10   &  {\rm from}  & P_3^{(\Sigma)}  
\end{array}   \;.  
\right. \label{treefit}
\end{equation}
We have used the experimental values listed in Table~\ref{results}, 
and the errors we quote for $\beta_{27}$ do not include any 
estimate of theoretical errors. The results in Eq.~(\ref{treefit}) 
are not inconsistent, but given the large experimental errors, 
it would be premature to say that the fit is good.

\begin{table}[ht]      
\caption{\label{results}%
Summary of results for  $\;|\Delta\bfm{I}|=3/2\;$  
components of the  S- and P-wave  amplitudes.   
We use the parameter values 
$\;\beta_{27}^{}=\delta_{27}^{}=\tilde{\beta}_{27}^{}=-0.068 \, 
\sqrt{2}\, f_{\!\pi}^{}G_{\rm F}^{} m_{\pi}^2\;$  as discussed in the text. 
In the Theory columns with the ``Octet'' and ``Decuplet'' headings,  
each $P_3^{}$ entry is written as a sum of two numbers, where  
the second one results from the baryon-mass renormalization in 
the tree-level P-wave diagrams.  
Each of the  $g_{27}^{}$  terms is also given as a sum of two numbers, 
the first one coming from diagrams with octet baryons only 
and the second one from graphs involving the decuplet.}
\centering   \small 
\vskip 1\baselineskip  
\begin{tabular}{crrrrrrl}    
\hline \hline      
&& \multicolumn{6}{c}{Theory}  
\\ \cline{3-8}    
Amplitude  &  Experiment\hspace{1ex}    &  
\raisebox{-1ex}{Tree\hspace{1ex}}       &  
\raisebox{-1ex}{Octet\hspace{1.5em}}    &   
\raisebox{-1ex}{Decuplet\hspace{2ex}}   &   
\raisebox{-1ex}{$g_{27}^{}$ term\hspace{2ex}}  &  
\raisebox{-1ex}{$\tilde{\beta}_{27}^{}$ term\hspace{-1.5ex}}   &
\\   
&&  
\raisebox{-1ex}{{\cal O}(1)\hspace{0.5ex}}         &   
\raisebox{-1ex}{${\cal O}(m_s^{}\ln{m_s^{}})$}     & 
\raisebox{-1ex}{${\cal O}(m_s^{}\ln{m_s^{}})$}     &   
\raisebox{-1ex}{${\cal O}(m_s^2 \ln{m_s^{}})$}     &   
\raisebox{-1ex}{${\cal O}(m_s^{})$\hspace{-1ex}}   &    
\smallskip \\ \hline      
\vspace{-2.5ex} && && && \\  
$S_3^{(\Lambda)}$ &  $-0.047\pm 0.017$    &   0\hspace{1em} &   
\hspace{5ex} 0\hspace{2.5em}   &  
\hspace{4ex} 0\hspace{2.5em}   &  
\hspace{2ex} 0.063$-$0.018     &   0.027    &
\\
$S_3^{(\Xi)}$          & $ 0.088\pm 0.020$      & 0\hspace{1em}    &  
0\hspace{2.5em}        & 0\hspace{2.5em}        & $-$0.051$-$0.033 & 
$-$0.036               &  
\\   
$S_3^{(\Sigma)}$       & $-0.107\pm 0.038$      & $-$0.107         &  
$-$0.089\hspace{1.5em} & $-$0.084\hspace{1.5em} & 0.003$+$0.079    &  
0.029                  &  
\\  
$P_3^{(\Lambda)}$      & $-0.021\pm 0.025$      & 0.012            &  
0.011$-$0.006          & $-$0.015$-$0.045       & 0.003$-$0.006    & 
0\hspace{1em}          &  
\\  
$P_3^{(\Xi)}$          & $ 0.022\pm 0.023$      & $-$0.037         &   
$-$0.055$+$0.031       & 0.046$+$0.019          & $-$0.001$-$0.000 &  
0\hspace{1em}          &  
\\  
\vspace{.3ex}   
$P_3^{(\Sigma)}$     & $-0.110\pm 0.045$    & 0.032            &   
0.031$-$0.016        & $-$0.047$-$0.124     & $-$0.003$+$0.002 &   
0\hspace{1em}        &  
\\  
\hline \hline 
\end{tabular}   
\vskip 1\baselineskip
\end{table}

Since the errors in the measurements of the P-wave amplitudes 
are larger than those in the S-wave amplitudes, we can take the 
point of view that we will fit the value of $\beta_{27}^{}$ 
to the S-wave $\Sigma$ decay and treat the P-wave amplitudes 
as predictions.  
Recall that in the analysis of the  $\;|\Delta\bfm{I}|=1/2\;$  
amplitudes, the two parameters that occur at lowest order in 
$\chi$PT theory are also extracted from a fit to S-wave amplitudes, 
and the P-wave amplitudes are then treated as predictions. 
In that case, the predictions for the P-wave amplitudes are 
completely wrong, differing from the measurements by factors of 
up to 30~\cite{bijnens,jenkins}. We show our results for the 
$\;|\Delta\bfm{I}|=3/2\;$  amplitudes in the column 
labeled ``Tree, ${\cal O}(1)$''  in Table~\ref{results}.   
These lowest-order predictions for the P-wave amplitudes
are not impressive, but they do have the right order 
of magnitude and they differ from the central value of 
the measurements by at most three standard deviations in the four cases 
where the predictions are non-zero.

At lowest order, two of the S-wave amplitudes 
vanish, $S_3^{(\Lambda)}$ and $S_3^{(\Xi)}$. This, of course, 
only indicates that these two amplitudes are predicted to be 
smaller than $S_3^{(\Sigma)}$ by about a factor of three since  
there are non-vanishing contributions from operators that occur at 
next order,  ${\cal O}(m_s^{}/\Lambda_{\chi SB}^{})$  
[for example, Eq.~(\ref{higheror})]. Again, this is not 
in conflict with experiment.

Beyond leading order in $\chi$PT,  there are too many unknown 
coefficients for the theory to be predictive. This makes it possible 
to fit all the amplitudes at next-to-leading order, but this 
fit is not particularly instructive. 
In this paper we limit ourselves to study the question of whether 
the lowest-order predictions are subject to large higher-order  
corrections.  
To address this issue, we look at our one-loop calculation of   
the  ${\cal O}(m_s^{}\ln{m_s^{}})$  corrections.   
In Table~\ref{results} we present these corrections in 
two columns, using a subtraction scale  $\;\mu = 1\,\rm GeV.\;$
The column marked  ``Octet, ${\cal O}(m_s^{}\ln{m_s^{}})$''  
is the result of those diagrams in which only octet baryons are 
allowed in the loops.  
(Contributions from the renormalization of the wave-function 
and decay constant of the pion are included in this column.)   
These results depend only on the constant  $\beta_{27}^{}$  and we   
use the value   
$\;\beta_{27}^{}=-0.068\,\sqrt{2}\, f_{\!\pi}^{}  
   G_{\rm F}^{} m_{\pi}^2\;$  
from the leading-order fit to  $S_3^{(\Sigma)}$ (of course they also
depend on the parameters from the strong sector, but these are 
already determined). 
For the P-waves we have split the results into the sum of two terms, 
the second one corresponding to baryon-mass renormalization in the
tree-level pole diagrams [the $\gamma_{BB^\prime}^{}$ of 
Eq.~(\ref{defpwave})] and the first one to everything else. We 
find it instructive to examine these two terms separately because, 
as we noted in Section~3, the second term was not included 
in the calculation of the $\;|\Delta\bfm{I}|=1/2\;$ amplitudes 
of Jenkins~\cite{jenkins}.\footnote{From our results, 
it appears that the inclusion of these terms could significantly 
affect the discussion of the  $\;|\Delta\bfm{I}|=1/2\;$  amplitudes. 
We will present our results for that case elsewhere.} 
We find that the $\gamma_{BB^\prime}^{}$  terms have 
very similar values to other terms of the same order. 
However, when combined with the other terms they can change 
the relative size of corrections to different amplitudes 
substantially (see also the column in Table~\ref{results} that 
shows the decuplet loops).

We can see in Table~\ref{results} that some of the loop corrections are 
as large as the lowest-order results even though they are expected to be 
smaller by about a factor of $M_K^2/(4\pi f_{\!\pi}^{})^2$. 
By studying  Eqs.~(\ref{octetsl})  and~(\ref{octetpl}),  
it is possible to see that 
each amplitude receives several contributions from different 
diagrams that combine to give the polynomials in $D$ and $F$. 
These terms add up constructively in some cases and nearly 
cancel out in others, resulting in deviations of up to an order 
of magnitude from the power counting expectation. 
This is an inherent flaw in a perturbative calculation where 
the expansion parameter is not sufficiently small.

The one-loop corrections are all much smaller 
than their counterparts in  $\;|\Delta\bfm{I}|=1/2\;$  
transitions, where they can be as large as 15 times 
the lowest-order amplitude in the case of the  P-wave in  
$\;\Sigma^+ \rightarrow n \pi^+\;$~\cite{jenkins}. 
In that case, Jenkins noted that this was due to an anomalously 
small lowest-order prediction arising from the cancellation of two 
nearly identical terms~\cite{jenkins}.

An important argument in deciding that 
the ${\cal O}(m_s^{}\ln{m_s^{}})$ corrections are a good estimate 
for the size of the complete next-to-leading-order 
corrections is that these are non-analytic terms that cannot 
arise from counterterms. In our calculation we find that there is 
another type of non-analytic correction at one-loop, of 
${\cal O}(\pi m_s^{3/2})$. These terms have the same origin as the 
terms proportional to $m_K^3$ that occur in the analysis 
of baryon masses~\cite{jenkins} that appear in Eq.~(\ref{mkcubed}).   
Numerically, we find that they are not 
important for the  $\;|\Delta\bfm{I}|=3/2\;$  transitions, and for 
this reason we do not discuss them in detail. 
They only affect $S_3^{(\Sigma)}$ where they decrease 
the size of the  ${\cal O}(m_s^{}\ln{m_s^{}})$ correction by about 10\%.

Next we consider the contributions of loop diagrams with 
decuplet baryons. It has been emphasized in Ref.~\cite{manjen} 
that the decuplet plays a special role in heavy-baryon $\chi$PT 
and that these terms can, therefore, be significant.   
Our results depend on one additional constant, $\delta_{27}^{}$,  
which cannot be fit from experiment because it does not 
contribute to any of the observed weak decays of 
a decuplet baryon. 
To illustrate the effect of these terms, we choose 
$\;\delta_{27}^{}=\beta_{27}^{},\;$  a choice consistent with 
dimensional analysis and with the normalization of  
Eq.~(\ref{loweak}).   
We present these results in the column marked 
``Decuplet, ${\cal O}(m_s^{}\ln{m_s^{}})$'' in Table~\ref{results}. 
For the P-waves, we have again split the contributions from 
$\gamma_{BB^\prime}^{}$  from everything else.   
The comments regarding the results for octet-baryon loops also 
apply here. 
We see that the decuplet-loop contributions could be important and  
occur at the same level as the octet-loop contributions.  
To some extent this result follows from our choice 
for $\delta_{27}^{}$, but it is illustrative of the special role of  
the decuplet in the formalism of Ref.~\cite{manjen}.   
Notice, for example, that the analogous parameters 
in the  $\;|\Delta\bfm{I}|=1/2\;$  sector  $h_D^{}$, $h_F^{}$ and  
$h_C^{}$ of  Eq.~(\ref{loweakoc})  end up being of the same order   
after they are fitted to the experimental amplitudes~\cite{jenkins}.
 
At this point in the calculation, with terms of order  
${\cal O}(1)$  and terms of order ${\cal O}(m_s^{}\ln{m_s^{}})$, 
we still find that $S_3^{(\Lambda)}$ and $S_3^{(\Xi)}$ are zero. 
We now discuss the non-zero contributions to these two 
amplitudes derived in Section~4.

In the column marked 
``$g_{27}^{}$ term, ${\cal O}(m_s^2\ln{m_s^{}})$''  
of  Table~\ref{results} we present the result from the
one-loop diagrams in which the weak transition occurs in a vertex 
that involves only mesons, Figure~\ref{g27,loop}. We present 
separate results for the contributions from loops with only 
octet baryons (first term) and loops with octet and decuplet baryons. 
The non-analytic part of these diagrams is uniquely determined 
in terms of $g_{27}^{}$ measured in  $\;K \rightarrow \pi\pi\;$  
decays and of couplings from the strong interaction Lagrangian. 
Since there can be contributions of  ${\cal O}(m_s^{})$   
from counterterms such as  Eq.~(\ref{higheror}),   
we cannot expect these terms to be the dominant ones.  
They do indicate, however, that $\chi$PT produces a non-zero  
$S_3^{(\Lambda)}$ and  $S_3^{(\Xi)}$ at the right level.

If we take, for example, $\;\tilde{\beta}_{27}^{}=\beta_{27}^{}\;$ 
in Eq.~(\ref{higheror}), we obtain terms of ${\cal O}(m_s^{})$ 
that are also of the same order as the measured  
$S_3^{(\Lambda)}$ and $S_3^{(\Xi)}$.   
We show these numbers in the column labeled   
``$\tilde{\beta}_{27}^{}$ term, ${\cal O}(m_s^{})$'' 
in Table~\ref{results}. 
We conclude that these two amplitudes can be accommodated 
naturally in $\chi$PT, but they cannot be predicted at present.

It is intriguing that some of the terms proportional to $g_{27}^{}$ 
are as large as they are in Table~\ref{results}.   
Formally they are of  ${\cal O}(m_s^2\ln{m_s^{}})$, and we have   
argued in Section~4 that we expect them to be smaller than the terms 
of  ${\cal O}(m_s^{}\ln{m_s^{}})$. 
For the same reason, we would also expect them to be smaller than 
the terms of  ${\cal O}(m_s^{})$  proportional 
to  $\tilde{\beta}_{27}^{}$. 
But a glance at Table~\ref{results} shows that this is not the case. 
A careful study of our results in Sections~3~and~4 indicates that 
if we compare the terms proportional to $g_{27}^{}$ with the 
terms proportional to $\beta_{27}^{}$, the former are ``suppressed'' 
by factors of order one times $m_s^{}/f_{\!\pi}^{}$.   
This explains the relative importance of these terms both in our 
calculation and in that of Jenkins \cite{jenkins}: although they are  
indeed proportional to an additional power of $m_s^{}$, the scale is  
set by  $f_{\!\pi}^{}$  instead of  $4\pi f_{\!\pi}^{}$.   
From studying the diagrams of Figure~\ref{g27,loop},  
it is obvious that there are no additional factors of $4\pi$   
associated with these terms relative to other one-loop terms.   
Also, since the weak transitions arise from the leading-order weak 
Lagrangian in the meson and baryon sectors, there are no relative 
factors of the scale of chiral symmetry breaking. The different 
relative importance of these terms in the different amplitudes 
is again due to the polynomials in $D$ and $F$. 
For example, the contribution of $g_{27}^{}$ terms to  
$S_3^{(\Sigma)}$ from baryon-octet loops looks relatively small.   
But if we calculate this number for the decay   
$\;\Sigma^+ \rightarrow n \pi^+,\;$   
we find that it is an order of magnitude larger. 
From our calculation, it is not possible to decide whether the  
$m_s^{}/f_{\!\pi}^{}$ factor is simply a numerical peculiarity of  
these diagrams, or whether it signals a breakdown of chiral  
perturbation theory for hyperon decays in the sense that there 
are some higher order terms where the scale is set 
by  $f_{\!\pi}^{}$.

For completeness, we summarize our results for arbitrary  
values of  $\beta_{27}^{}$,  $\delta_{27}^{}$  and  
$\tilde{\beta}_{27}^{}$  in  Table~\ref{resultsgen}.  
The terms proportional to  $g_{27}^{}$  are the same as in 
Table~\ref{results} since this constant is fixed.

Finally, it is worth pointing out that it is possible to obtain  
a good fit to experiment with just the terms considered in  
this paper. 
A least-squares fit to the four amplitudes that are not zero 
at  ${\cal O}(m_s^{}\ln{m_s^{}})$  yields  $\;\chi^2 \approx 1.2,\;$  
and the extracted values of  $\beta_{27}^{}$  and  $\delta_{27}^{}$   
lead to the theoretical numbers shown in Table~\ref{resultsfit}.  
Similarly, a one-parameter fit to  $S_3^{(\Lambda)}$  and   
$S_3^{(\Xi)}$  has  $\;\chi^2\approx 2.8,\;$ 
and the result is also shown in Table~\ref{resultsfit}.

In conclusion, we have presented a discussion of  
$\;|\Delta\bfm{I}|=3/2\;$  amplitudes for hyperon non-leptonic 
decays in $\chi$PT. 
At leading order these amplitudes are described in terms of 
only one parameter. 
This parameter can be fit from the observed value of 
the S-wave amplitudes in $\Sigma$ decays.  
After fitting this number, we have made certain predictions for 
the other amplitudes, some quantitative (the P-waves) some 
qualitative (the other S-waves). 
We have used our one-loop calculation to discuss 
the robustness of the predictions. 
Our predictions are not contradicted by current data, 
but current experimental errors are too large for 
a meaningful conclusion. We have shown that the one-loop 
non-analytic corrections have the relative size expected from 
naive power counting. 
The combined efforts of E871 and KTeV could give us improved 
accuracy in the measurements of some of the decay modes that 
we have discussed and allow a more quantitative comparison of 
theory and experiment.
\begin{table}[hb]      
\caption{\label{resultsgen}%
$\;|\Delta\bfm{I}|= 3/2\;$  components of the  S- and P-wave  
theoretical amplitudes.  
Only contributions proportional to $\beta_{27}^{}$,  
$\delta_{27}^{}$  or  $\tilde{\beta}_{27}^{}$ are tabulated here.}  
\centering   \small 
\vskip 1\baselineskip 
\begin{tabular}{crrrlrrrl}    
\hline \hline      
\raisebox{-2ex}{Amplitude}   &  
\raisebox{-2ex}{$\begin{array}{c} \displaystyle  
{\rm Tree}  \\ {\cal O}(1)   
\end{array}$}\hspace{1.5ex}   
&   
\multicolumn{3}{c}{\raisebox{-2ex}{$\begin{array}{c} \displaystyle  
{\rm Octet}  \\ {\cal O}(m_s^{}\ln{m_s^{}})
\end{array}$}}
&  
\raisebox{-2ex}{$\begin{array}{c} \displaystyle  
{\rm Decuplet}  \\ {\cal O}(m_s^{}\ln{m_s^{}})
\end{array}$}\hspace{4ex}   
&   
\multicolumn{3}{c}{\raisebox{-2ex}{$\begin{array}{c} \displaystyle  
\tilde{\beta}_{27}^{} \;{\rm term} \\ {\cal O}(m_s^{})
\end{array}$}\hspace{2ex}}   
\smallskip \\ \hline      
\vspace{-2.5ex} && && && & \\  
$S_3^{(\Lambda)}$  &   $ 0\hspace{2em}$   &&   $0\hspace{2em}$   &&   
                       $ 0\hspace{4em}$   &&   
                       $-0.399\,\tilde{\beta}_{27}^{}$  &  
\\   
$S_3^{(\Xi)}$      &   $ 0\hspace{2em}$   &&  $0\hspace{2em}$   &&   
                       $ 0\hspace{4em}$   &&   
                       $ 0.528\,\tilde{\beta}_{27}^{}$  &  
\\   
$S_3^{(\Sigma)}$   &   $ 1.581\,\beta_{27}^{}$   &&   
                       $ 1.316\,\beta_{27}^{}$   &&   
                       $ 1.466\,\beta_{27}^{}
                        -0.230\,\delta_{27}^{}$   &&     
                       $-0.428\,\tilde{\beta}_{27}^{}$  &  
\\   
$P_3^{(\Lambda)}$  &   $-0.174\,\beta_{27}^{}$   &&   
                       $-0.072\,\beta_{27}^{}$   &&   
                       $ 0.870\,\beta_{27}^{}   
                        +0.025\,\delta_{27}^{} $   &&   
                       $ 0\hspace{2em}$   & 
\\        
$P_3^{(\Xi)}$      &   $ 0.546\,\beta_{27}^{}$   &&   
                       $ 0.359\,\beta_{27}^{}$   &&   
                       $-0.610\,\beta_{27}^{}  
                        -0.358\,\delta_{27}^{}$   &&   
                       $ 0\hspace{2em}$   & 
\\  
\vspace{.3ex} 
$P_3^{(\Sigma)}$   &   $-0.475\,\beta_{27}^{}$   &&   
                       $-0.223\,\beta_{27}^{} $   &&   
                       $ 2.466\,\beta_{27}^{}  
                        +0.069\,\delta_{27}^{}$   &&   
                       $ 0\hspace{2em}$   & 
\\  
\hline \hline 
\end{tabular}   
\vskip 1\baselineskip
\end{table}
\begin{table}[ht]      
\caption{\label{resultsfit}%
Results of least-squares fits to  the $\;|\Delta\bfm{I}|=3/2\;$  
components of the  S- and P-wave  amplitudes,  
compared with experiment.  
The parameter values preferred by the fits are 
$\;\beta_{27}^{}=-0.033,\;$  $\delta_{27}^{}=-0.103,\;$  and  
$\;\tilde{\beta}_{27}^{}=0.283,\;$  all in units of    
$\;\sqrt{2}\, f_{\!\pi}^{}G_{\rm F}^{} m_{\pi}^2.\;$} 
\centering   \small 
\vskip 1\baselineskip  
\begin{tabular}{crr}    
\hline \hline      
\raisebox{-1ex}{Amplitude}  &   
\raisebox{-1ex}{Experiment\hspace{1ex}}    &   
\raisebox{-1ex}{Theory\hspace{-1ex}}   
\smallskip \\ \hline      
\vspace{-2.5ex} && \\  
$S_3^{(\Lambda)}$      &  $-0.047\pm 0.017$     & $-$0.068
\\  
$S_3^{(\Xi)}$          & $ 0.088\pm 0.020$      & 0.066
\\   
$S_3^{(\Sigma)}$       & $-0.107\pm 0.038$      & $-$0.120         
\\  
$P_3^{(\Lambda)}$      & $-0.021\pm 0.025$      & $-$0.023            
\\  
$P_3^{(\Xi)}$          & $ 0.022\pm 0.023$      & 0.027         
\\  
\vspace{.3ex}   
$P_3^{(\Sigma)}$       & $-0.110\pm 0.045$      & $-$0.066            
\\  
\hline \hline 
\end{tabular}   
\vskip 1\baselineskip
\end{table}

\vspace{1in}

\noindent {\bf Acknowledgments} This work  was supported in
part by DOE under contract number DE-FG02-92ER40730. The work of 
A.~El-Hady was supported in part by DOE under contract number 
DE-FG02-87ER40371. G.~V. thanks the Special Center for the 
Subatomic Structure of Matter at the University of Adelaide for 
their hospitality while part of this work was done. We thank 
John F.~Donoghue, Xiao-Gang~He, K.~B.~Luk and Sandip Pakvasa
for helpful discussions.

\end{document}